\documentclass{emulateapj}
\usepackage{rotating}
\shorttitle{Comparing NS Kicks to SNR Asymmetries}
\shortauthors{Holland-Ashford et al.} 

\newcommand{\ltsima}{$\; \buildrel < \over \sim \;$}
\newcommand{\simlt}{\lower.5ex\hbox{\ltsima}}

\newcommand{\ls}{{_<\atop^{\sim}}}

\newcommand{\gs}{{_>\atop^{\sim}}}

\def\arcmin{\hbox{$^\prime$}}
\def\arcsec{\hbox{$^{\prime\prime}$}}

\begin{document}

\title{Comparing Neutron Star Kicks to Supernova Remnant Asymmetries}

\author{Tyler Holland-Ashford\altaffilmark{1}$^,$\altaffilmark{2}, 
Laura A. Lopez\altaffilmark{1}$^,$\altaffilmark{2}, 
Katie Auchettl\altaffilmark{2}$^,
$\altaffilmark{3},
Tea Temim\altaffilmark{4},
Enrico Ramirez-Ruiz\altaffilmark{5}$^,$\altaffilmark{6}}
\altaffiltext{1}{The Ohio State University Department of Astronomy, 140 W 18th Ave, Columbus, OH 43201, USA}
\altaffiltext{2}{The Ohio State University Center for Cosmology and Astro-particle Physics, 191 West Woodruff Ave, Columbus, OH 43210, USA}
\altaffiltext{3}{Department of Physics, The Ohio State University, 191 W. Woodruff Avenue, Columbus, OH 43210, USA}
\altaffiltext{4}{Space Telescope Science Institute, 3700 San Martin Drive, Baltimore, MD 21218, USA}
\altaffiltext{5}{Department of Astronomy \& Astrophysics, University of California, Santa Cruz, CA 95064, USA}
\altaffiltext{6}{Niels Bohr Institute, University of Copenhagen, Copenhagen 2100, Denmark}

\email{holland-ashford.1@osu.edu}  

\begin{abstract}

Supernova explosions are inherently asymmetric and can accelerate new-born neutron stars (NSs) to hundreds of km s$^{-1}$. Two prevailing theories to explain NS kicks are ejecta asymmetries (e.g., conservation of momentum between NS and ejecta) and anisotropic neutrino emission. Observations of supernova remnants (SNRs) can give us insights into the mechanism that generates these NS kicks. In this paper, we investigate the relationship between NS kick velocities and the X-ray morphologies of 18 SNRs observed with the {\it Chandra} X-ray Observatory and the {\it R$\ddot{\rm o}$ntgen} Satellite ({\it ROSAT}). We measure SNR asymmetries using the power-ratio method (a multipole expansion technique), focusing on the dipole, quadrupole, and octupole power-ratios. Our results show no correlation between the magnitude of the power-ratios and NS kick velocities, but we find that for Cas~A and G292.0$+$1.8, whose emission traces the ejecta distribution, their NSs are preferentially moving opposite to the bulk of the X-ray emission. In addition, we find a similar result for PKS 1209--51, CTB~109, and Puppis~A; however their emission is dominated by circumstellar/interstellar material, so their asymmetries may not reflect their ejecta distributions. Our results are consistent with the theory that NS kicks are a consequence of ejecta asymmetries as opposed to anisotropic neutrino emission. In the future, additional observations to measure NS proper motions within ejecta-dominated SNRs are necessary to constrain robustly the NS kick mechanism.
\end{abstract} 


\section{Introduction}
In the past decades, evidence has mounted that supernova explosions (SNe) can have significant deviations from spherical symmetry. Spectropolarimetric studies that measure the polarization of light 
as it is scattered through the debris layers of expanding SNe have demonstrated that both Type Ia and core-collapse (CC) SNe are aspherical near maximum brightness (e.g., \citealt{wang08,kasen09,inserra16}). Line profiles in nebular spectra (100--200 days after explosion) of SNe also show evidence of ejecta asymmetries (e.g., \citealt{mazzali01,maeda10,maeda12,uchida13,winkler14}). Confirmation of these asymmetries is also possible via comparison of the SN light echo spectra from several dust concentrations around SNRs (e.g., \citealt{rest11}). 

Neutron stars (NSs), the compact objects formed in some CC SNe, may provide information about the SN explosion mechanism and its effect on ejecta asymmetries. NSs are typically `kicked' in the explosions to velocities of hundreds of km s$^{-1}$ \citep{lyne94,arz02,hobbs05,faucher06}, with some sources reaching velocities of $\sim$ 1000 km s$^{-1}$ (e.g., \citealt{chatterjee05,winkler07}). These 1000 km s$^{-1}$ velocities are larger than can be accounted for by the disruption of a close binary system (which typically produce velocities of $\sim$100 km s$^{-1}$: \citealt{lai01}), suggesting that the explosion is the likely cause of these kicks.

Two theories have been proposed to explain the cause and direction of NS kicks. In the first scenario (hereafter the ejecta asymmetry model), hydrodynamical instabilities lead to asymmetric mass ejection, accelerating the NS in a direction opposite to the bulk of ejecta \citep{scheck06, wongwathanarat13,janka17}. For example, simulations by \cite{wongwathanarat13} reveal that gravitational forces from the anisotropic ejecta can generate NS recoil velocities of at least $\sim$700 km s$^{-1}$, and 2D models by \cite{scheck06} have achieved NS velocities exceeding 1000 km s$^{-1}$. This model predicts that the heavy elements (e.g., iron, titanium) are expelled in a direction opposite to the NS kick, while the intermediate-mass elements (e.g., carbon, oxygen, neon, magnesium) are only marginally affected.

In the second scenario, anisotropic neutrino emission carries away the bulk of the gravitational binding energy of the neutron star, and both the NS and the ejecta move opposite to the direction of the majority of that energy \citep{fryer06}. However, several authors have noted that strong ($\gs$10$^{16}$ G) magnetic fields and specific assumptions about the field configuration or nonstandard neutrino physics are necessary to generate kick velocities of $\sim$300 km s$^{-1}$ \citep{scheck06,wongwathanarat13}. Both of these models can reproduce the observed spin-kick alignment in pulsars \citep{spruit98,johnston05,fryer06,janka17,muller17}.

Supernova remnants (SNRs) offer the means to distinguish between these two scenarios. Several hundred SNRs have been identified in the Milky Way and the Large and Small Magellanic Clouds (e.g., \citealt{badenes10,green14}), and many of these sources have detected compact objects and/or pulsar wind nebulae (e.g., \citealt{gaensler06,deluca08}). The metals synthesized in the explosions are shock-heated to X-ray emitting temperatures, and X-ray observations of SNRs have revealed complex morphologies that may reflect asymmetries in SN explosion mechanisms (see \citealt{weiss06b} and \citealt{vink12} for a review). For example, the young SNR Cassiopeia~A (Cas~A) has asymmetric velocity gradients in its ejecta knots and jets that are attributed to its explosion \citep{delaney10,fesen06,hwang12,milisavljevic15,grefenstette17}. 

The complex structure of SNRs makes it difficult to systematically compare sources. To address this challenge, \cite{lopez09a} developed techniques to quantitatively characterize emission from SNRs. Subsequent work showed that the X-ray and infrared morphologies of CC SNRs are more elliptical and mirror asymmetric than those of Type Ia SNRs \citep{lopez09b,lopez11,peters13}. The origin of these asymmetries remains unclear: they may reflect the asymmetries inherent to the explosion mechanism, or they may arise from interactions with an inhomogeneous medium. For example, kinematic studies demonstrate large-scale asymmetries in SNR ejecta (e.g., SN~1987A: \citealt{boggs15}; Cas~A: \citealt{rest11,grefenstette17}), while observations and simulations show that interaction with a dense medium can alter the morphology and thermodynamic properties of SNRs (e.g., \citealt{tenorio85,lazendic06,slane15}).

In this paper, we examine the relationship between the X-ray morphology of a sample of Milky Way SNRs and their NS velocities to assess both the role of the explosion mechanism in shaping SNRs and to probe the origin of NS kicks. For this work, we exploit multi-epoch {\it Chandra} X-ray Observatory archival images to measure NS transverse velocities, and we compare them to the asymmetries of the X-ray morphologies as observed by {\it Chandra} and the {\it R$\ddot{\rm o}$ntgen} Satellite ({\it ROSAT}). In Section~\ref{sec:data}, we describe the observations used in this study and the selection criteria to build our sample. In Section~\ref{sec:methods}, we explain the methods used to analyze the X-ray morphology and to measure the NS velocities. In Section~\ref{sec:results}, we present our results, while in Section~\ref{sec:disc}, we discuss the implications regarding the origin of NS kicks. Section 6 summarizes our conclusions and outlines possible future work that would test SN models using SNRs and their compact objects.

\section{Observations, Sample, and Data Analysis} \label{sec:data}

  \begin{figure*}
  \includegraphics[width=\textwidth]{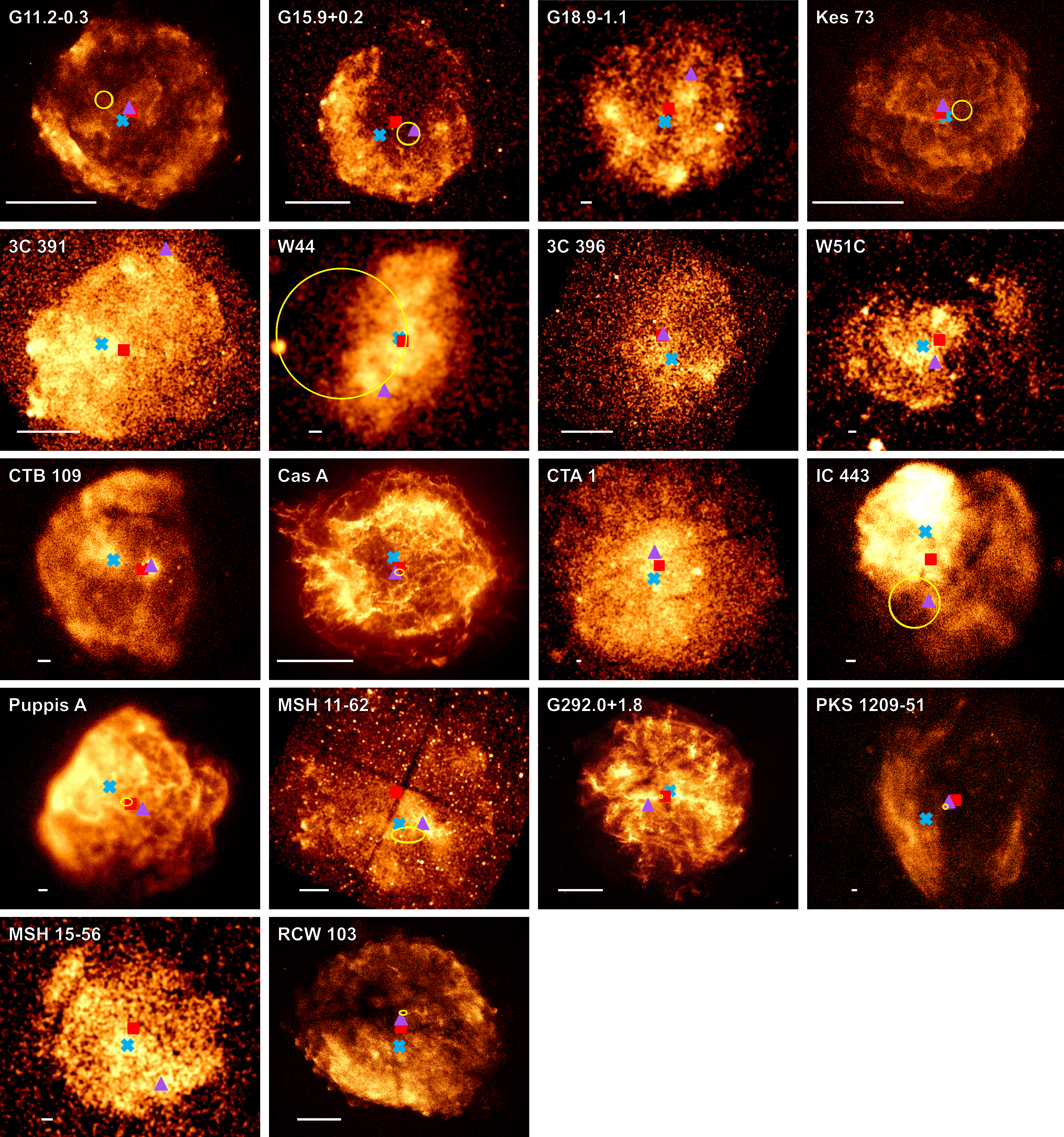}

 \caption{Images of the SNR sample before point source removal in the order they are listed in Table~\ref{table:SNRtable}. Blue crosses are the calculated centers-of-emission, red squares are the geometric centers, and purple triangles denote the location of the NS associated with each SNR. The yellow circles are the 1-$\sigma$ confidence interval for the explosion site determined via back-evolving the NS's proper motion, except in Cas~A and G292.0$+$1.8 where the explosion sites are determined from back-evolving filament motions. Only the SNRs with measured proper motions from multi-epoch imaging have yellow circles (see Tables~\ref{table:obslog} and~\ref{table:SNRtable}). The white scale bars are 2 arcminutes. North is up, and east is left.}
  \label{fig:snrs} 
 \end{figure*}

We utilize archival X-ray observations of 18 SNRs selected based on the criteria discussed in the following paragraphs. Specifically, we employ X-ray imaging data from the {\it Chandra} Advanced CCD Imaging Spectrometer (ACIS) and {\it ROSAT}'s Position Sensitive Proportional Counter (PSPC). We did not include {\it XMM-Newton} observations in our analysis because no targets that met our selection criteria were imaged only by {\it XMM-Newton}. One source, CTB~109, has been observed by both {\it ROSAT} and {\it XMM-Newton}, and we opted to use the former because the SNR was imaged fully in one {\it ROSAT} pointing, whereas the {\it XMM-Newton} mosaic required stitching of several pointings together.

Images of the SNRs are shown in Figure~\ref{fig:snrs}. Our sample is comprised of relatively young ($\ls$20~kyr old) Milky Way CC SNRs that are imaged fully by {\it Chandra} or {\it ROSAT}, have strong thermal X-ray emission, and have a detected NS with a reported or measurable transverse velocity. SNRs are typically classified as CC SNRs based on their elemental abundances and on the presence of a neutron star near their centers (see references in Table~\ref{table:SNRtable}). We do not consider SNRs in other galaxies because they are too distant to measure NS proper motions. In addition, we exclude from our sample the SNRs dominated by non-thermal emission, as the X-ray morphologies would reflect the structure of the synchrotron radiation from the forward shock or a large PWN. For example, G21.5--0.9 was excluded because non-thermal emission from the PWN covers a large fraction of the remnant and dominates the full X-ray band \citep{safi01}. 

Since we are investigating the spatial distribution of thermal X-ray emission, we only selected remnants whose emission is dominated ($\gs$90\%) by thermal emission in the 0.5--2.1 energy band. For the SNRs imaged fully by {\it Chandra}, we extracted X-ray spectra from multiple regions and fit the data in each region using XSPEC Version 12.9.0 \citep{arnaud96} with AtomDB\footnote{atomdb.org} 3.0.7 \citep{smith01,foster12}. For the larger SNRs imaged by {\it ROSAT}, we extracted and fit the X-ray spectra from multiple regions of {\it Chandra} observations if available( see Table~\ref{table:obslog}). By fitting multiple regions, it allowed us to account for variations in the thermal properties across the remnant. We modeled all of the data as an absorbed ({\it phabs}) thermal plasma  ({\it vapec} or {\it vpshock}) plus a power-law component.  We allowed the elemental abundances (e.g., Mg, Si, S, Ar, Fe) to vary and capped the photon index to below 3.0 so the power-law component would reflect non-thermal emission. We used the solar abundance table from \cite{anders89}. We adopted the best-fit results to compute the total unabsorbed flux contributed by the thermal and power-law components. In Table~\ref{table:SNRtable} we have listed the relevant properties, including estimated ages, distances, and neutron star velocities (derived in this work or found in the literature) for our sample of 18 SNRs. 

Data reduction and analysis was performed using the {\it Chandra} Interactive Analysis of Observations ({\sc ciao}) Version 4.7 and {\sc ftools} \citep{blackburn95}. For the remnants with diameters smaller than 16\arcmin\ in X-rays, we produced exposure-corrected {\it Chandra} images in the 0.5--2.1 keV band, merging observations via the {\sc ciao} command {\it merge\_obs}. We then removed point sources using the commands {\it wavdetect} and {\it dmfilth}, including removal of the NS and associated PWN (if applicable) from the images. For all SNRs except for CTB 109, the PWN size is $\ls$2\% of the radius of the remnant, and thus their removal does not dramatically affect the morphology of the SNRs. The PWN in CTB~109 is approximately 10\% of the SNR radius. However, since the power-ratios weight more heavily emission at large radii, removal of the PWN near the origin has minimal effect on their values.

For the larger remnants, we analyzed {\it ROSAT} images but did not exposure correct them as doing so produced artifacts in the images that affected the power-ratio results. For example, the power-ratio values for G11.2--0.3 differed between the {\it Chandra} and exposure-corrected {\it ROSAT} images, but the values were consistent when using exposure-uncorrected {\it ROSAT} images. This effect held true for all SNRs with both {\it Chandra} and {\it ROSAT} observations, and thus we opted against exposure correction for the {\it ROSAT} images to enable comparison between the {\it Chandra} and {\it ROSAT} results. 

\subsection{Properties of the SNRs} \label{sec:csm}

As we aim to compare the ejecta distribution to the NS kick velocities, it is important to consider the contribution of ejecta and circumstellar/interstellar material (CSM/ISM) to the thermal X-ray emission of each remnant. Generally, the components that contribute most to the  thermal X-ray emission in SNRs are bremsstrahlung continuum and emission lines from collisionally-ionized plasmas. Depending on the age and ambient density, the plasmas may be under-ionized (in non-equilibrium ionization: NEI) or in collisional ionization equilibrium (CIE; we note that ionization timescales $\gs$10$^{12}$ s cm$^{-3}$ suggest the plasma is in CIE: \citealt{smith10}). Alternatively, some SNRs have evidence of recombining and over-ionized plasmas thought to arise from rapid cooling caused by molecular cloud interactions (e.g., see \citealt{kawasaki05}). Given the low densities and young ages of our sample, thermal emission can arise from either the shock-heated ejecta or the CSM/ISM. 

In the following paragraphs, we summarize the contribution of ejecta and CSM/ISM in each source's X-ray emission based on previous studies. Ejecta emission is characterized by super-solar abundances of metals such as Mg, Si, and S, and CSM/ISM emission has sub-solar or solar metal abundances. In sources where the emission is primarily from ejecta, the asymmetries measured by the power-ratio method reflect the ejecta distribution from the explosion. For SNRs that have significant emission from CSM/ISM or dense molecular clouds, the asymmetries measured by the power-ratio method are influenced by the environment and thus may not reflect the asymmetries of the explosion. In our sample, multiple remnants show evidence of interactions with CSM/ISM (3C~391, W44, 3C~396, W51C, CTB~109, IC~443, Puppis~A, and RCW~103; see details within this section), and we mark them as interacting (see e.g, Figure~\ref{fig:com}). 

{\bf G11.2$-$0.3}--G11.2$-$0.3 was first detected at X-ray wavelengths with {\it Einstein} HRI \citep{downes84} and was subsequently observed with {\it ROSAT} \citep{reynolds94} and with {\it Chandra} \citep{kaspi01,roberts03,lopez11,borkowski16}. The PWN and 65-ms period pulsar associated with G11.2$-$0.3 were discovered using {\it ASCA} observations \citep{vasisht96,torii97}, and \cite{kaspi01} showed that the pulsar is only 8\arcsec\ from the geometric center of the SNR. \cite{borkowski16} used multiple epochs of {\it Chandra} observations to measure a shell expansion rate that implied an age of 1400--2400 years, making G11.2$-$0.3 one of the youngest identified CC SNRs in the Milky Way. \cite{lopez11} performed a spatially-resolved spectroscopic analysis of 23 regions using {\it Chandra} data and found that the spectra were best-fit by a single absorbed, NEI component with absorbing columns of \hbox{$N_{\rm H}\approx(2.1-2.9)\times10^{22}$~cm$^{-2}$}, temperatures of $kT\approx0.6-1.4$~keV, and ionization timescales of $n_{\rm e}t\approx (0.9-10)\times10^{11}$ s cm$^{-3}$. The {\it Chandra} X-ray spectra have prominent Mg, Si, and S lines, and the fits by \cite{lopez11} showed these elements have supersolar abundances in most regions, indicating an ejecta origin of the emission.

{\bf G15.9$+$0.2}--G15.9$+$0.2 is a young (a few thousand year old) SNR with a shell-like morphology in X-rays \citep{reynolds06}. Using deep {\it Chandra} observations, \cite{klochkov16} showed that the X-ray point source CXOU J181852.0$-$150213 in G15.9$+$0.2 is a cooling, low-magnetized neutron star of the central compact object (CCO) class (e.g., \citealt{pavlov04}). Recently, \cite{maggi17} analyzed archival {\it XMM-Newton} observations of G15.9$+$0.2, extracting spectra from several locations along the shell and interior. These data have strong Mg, Si, S, Ar, and Ca lines, as well as a weak Fe K feature. \cite{maggi17} found that the spectra were best-fit by an absorbed, NEI plasma model, with absorbing columns of $N_{\rm H} \approx(4-5)\times10^{22}$ cm$^{-2}$, temperatures of $kT \approx 0.7-1.5$~keV, and ionization timescales of $n_{\rm e}t\approx(7-9)\times10^{10}$ s cm$^{-3}$ (except in the northwest, where the faint shell has $n_{\rm e} t > 3.5\times10^{11}$ s cm$^{-3}$). All regions showed enhanced abundances (with $\approx1.3-4.3\times$ the solar values) of the detected metals. 

{\bf G18.9$-$1.1}--G18.9$-$1.1 is a composite SNR that was first observed at X-ray wavelengths with a pointed {\it ROSAT} observation by \cite{fuerst97}. They found that the spectra were best described by an absorbed, NEI plasma model with temperature $kT \approx 0.95$~keV and a column density of $N_{\rm H} = (3.4\pm1.5)\times10^{21}$ cm$^{-2}$. Subsequently, \cite{harrus04} analyzed {\it ROSAT} and {\it ASCA} observations of G18.9$-$1.1. These authors showed that the spectra from the SNR were predominantly thermal in nature, and the best-fit model was an absorbed, thermal plasma in non-equilibrium ionization, with $N_{\rm H} \approx 10^{22}$ cm$^{-2}$, a temperature of $kT \approx 0.9$~keV, and an ionization timescale of $n_{\rm e} t \approx 1.1\times10^{10}$ s cm$^{-3}$. \cite{harrus04} found marginal evidence for supersolar abundances of Mg and Si in their fits. Using follow-up {\it Chandra} observations of G18.9$-$1.1, \cite{tullmann10} identified a faint X-ray point source, CXOU J182913.1$-$125113, with an associated trail of diffuse emission thought to be a PWN. Spectra extracted from the region of the PWN required a power-law component as well as a thermal component to adequately fit the data, and \cite{tullmann10} interpreted this result as evidence of contamination from the SNR's thermal emission.

{\bf Kes~73}--Kes~73 (G27.4$+$0.0) is a young ($\ls$2000 years old), shell-type SNR with the magnetar 1E 1841$-$045 located near its center \citep{gotthelf97,vasisht97}. Due to its small angular extent ($\approx$4\arcmin\ in diameter), {\it Chandra} and {\it XMM-Newton} observations were necessary to disentangle the magnetar's emission from the ejecta; the latter has a clumpy substructure throughout the remnant \citep{lopez11,kumar14}. \cite{kumar14} performed fits to the {\it Chandra} and {\it XMM-Newton} X-ray spectra from nine regions of Kes~73 and found that the data were best described by two absorbed, thermal components, one hot NEI plasma (with $kT \approx 1.6$~keV and $n_{\rm e} t \approx[0.5-3]\times10^{11}$ s cm$^{-3}$) with solar abundances, and one colder component (with $kT \approx0.5$~keV) near CIE (with $n_{\rm e} t \gs6\times10^{11}$ s cm$^{-3}$), with supersolar abundances of Si and S. \cite{lopez11} extracted {\it Chandra} spectra from 30 clumps identified in X-ray images, and their data were best-fit by an absorbed, single thermal plasma in NEI, with absorbing columns of $N_{\rm H} \approx(2-3)\times10^{22}$ cm$^{-2}$, temperatures of $kT \approx 0.6-1.2$~keV, and ionization timescales of $n_{\rm e} t \approx(0.8-45)\times10^{11}$ s cm$^{-3}$. These authors found supersolar abundances of Mg, Si, and S in most regions, indicating an ejecta origin of the thermal emission.

{\bf 3C~391 (G31.9$+$0.0)}--3C~391 was first detected in X-rays using {\it Einstein} \citep{wang84} and has been extensively followed up in multiple wavelengths. Radio observations revealed that the SNR exhibits a shell-like morphology, while extended radio emission indicative of a shock breakout into a low density environment is seen towards the southeast \citep{1993AJ....105.2226R}. Two 1720 MHz OH masers \citep{1996AJ....111.1651F} as well as enhanced CO \citep{1998AJ....115..247W} and [O\textsc{i}] \citep{1996A&A...315L.277R} emission is seen along the shell, evidence that the remnant is interacting with a nearby molecular cloud \citep[e.g.,][]{1999ApJ...511..836R}. 3C~391 has been studied in X-rays using {\it Einstein} \citep{wang84}, {\it ROSAT} \citep{1996ApJ...467..698R}, {\it ASCA} \citep{2001ApJ...563..202C}, {\it Chandra} \citep{chen04} and {\it Suzaku} \citep{sato14}. These observations revealed a center-filled morphology characteristic of mixed-morphology SNRs \citep{1998ApJ...503L.167R}. Using a deep {\it Chandra} observation, \citet{chen04} found that its X-ray emission is best described using a NEI plasma with approximately solar abundances of Mg, Si and S, and a uniform temperature throughout the SNR of $kT\approx0.2$~keV approaching CIE. Follow-up observations using {\it Suzaku} further confirmed the presence of a near-CIE plasma; however, \citet{sato14} found that the X-ray spectrum  of 3C~391 requires an additional high-temperature ($kT\approx0.50$~keV) recombining plasma (RP) component. The detection of enhanced Ca led \citet{sato14} to suggest that the RP component arises from ejecta from a 15M$_{\odot}$ progenitor, and its high ionization timescale ($n_{\rm e}t\approx 10^{12}$ s cm$^{-3}$) makes it one of the only SNR that shows evidence of overionization whose plasma is in/close to ionization equilibrium.

{\bf W44 (G34.7$-$0.4)}--Due to its center-filled X-ray morphology and bright radio synchrotron shell, W44 is one of the most well studied mixed-morphology SNRs in radio \citep[e.g.,][]{1972A&A....20..237K, 1987PASJ...39..709H, 1993MNRAS.265..631J}, X-rays \citep[e.g,][]{harrus97, 1994ApJ...430..757R, shelton04, kawasaki05, uchida12} and gamma-rays \citep[e.g.,][]{2013Sci...339..807A}. This 20 kyr \citep{1991ApJ...372L..99W} remnant shows strong evidence of  molecular cloud/shock interaction \citep[e.g.,][]{1997ApJ...489..143C, 1998ApJ...508..690F, 1998ApJ...505..286S, 2004AJ....127.1098S, 2005ApJ...618..297R, uchida12}, and it is one of the brightest GeV gamma-ray SNRs detected \citep[e.g.,][]{2013Sci...339..807A}. \cite{shelton04} found metal-rich ejecta at the center of the SNR using {\it Chandra} observations of W44. \cite{uchida12} confirmed these results with {\it Suzaku} and showed that the X-ray spectra of W44 are best described as an overionized plasma with enhanced abundances of Ne, Mg, Si, S, Ar, and Ca. There is a slight temperature gradient in the remnant, with a hotter temperature ($kT\approx0.48$~keV) at its center compared to its outer regions ($kT\approx0.40$~keV). \citet{shelton04} suggested that the complicated morphology and nature of this remnant arises from either evaporation of swept-up clouds \citep{1991ApJ...373..543W} or from thermal conduction \citep{1999ApJ...524..179C, 1999ApJ...524..192S}. The hard X-rays are not associated with the pulsar (B1953+01: \citealt{1991ApJ...372L..99W}), so the emission may arise from particles accelerated by the SNR shock front.

{\bf 3C~396 (G39.2$-$0.3)}--3C~396 is a middle-aged ($\approx$3000--7000~years old), composite SNR \citep{harrus99,su11}. {\it ASCA} observations of 3C~396 revealed a non-thermal component dominates at the center of 3C~396 \citep{harrus99}, and {\it Chandra} observations resolved this emission as a PWN \citep{olbert03}. \cite{su11} extracted {\it Chandra} X-ray spectra from multiple locations around the perimeter of 3C~396, and emission lines from Si, S, Ar, and Ca were evident. They found the data were best-fit by an absorbed, single NEI plasma, with absorbing columns of $N_{\rm H} \approx (4.4-5.5)\times10^{22}$ cm$^{-2}$, temperatures of $kT \approx 0.7-1.3$~keV, and ionization timescales of $n_{\rm e}t \approx(0.4-2.7)\times10^{11}$ s cm$^{-3}$ (although the southwest and east of the SNR only had lower-limits on $n_{\rm e}t$ of $>4\times10^{11}$ s cm$^{-3}$ and $>0.8\times10^{11}$ s cm$^{-3}$, respectively). \cite{su11} obtained supersolar abundances of Si, S, and Ca in several regions of the SNR, indicative of an ejecta origin for the thermal emission despite the moderate age of 3C~396. We note that 3C~396 is thought to be interacting with molecular clouds based on $^{12}$CO $J = 1-0$ and $^{12}$CO $J = 2-1$ observations, particularly on the western side where the SNR appears to be confined by a molecular wall \citep{su11}.

{\bf W51C (G49.2$-$0.7)}--W51C is located in a complicated region of the sky and is associated with a number of compact H{\sc ii} regions, dense molecular material, and the massive star forming region W51B \citep[e.g.,][]{1975LNP....42..443B, 1979ApJ...232..451M, 1997ApJ...475..194K,  1998AJ....116.1856C, 1999ApJ...518..760K, 2004MNRAS.353.1025K}. W51C has been detected in radio \citep{1991MNRAS.250..127C, 1995MNRAS.275..755S} all the way to TeV energies \citep[e.g.,]{ 2009ApJ...700L.127A, 2009AIPC.1112...54F, 2012A&A...541A..13A, 2016ApJ...816..100J} It shows evidence of interaction with nearby molecular clouds based on the detection of two OH masers \citep{1997AJ....114.2058G, 2013ApJ...771...91B}. This remnant has been extensively studied in X-rays using { \it Einstein} \citep{1990ApJS...73..781S}, {\it ROSAT} \citep{koo95}, {\it ASCA} \citep{2002AJ....123.1629K}, {\it Chandra} \citep{koo05, 2013ApJ...771...91B}, {\it XMM-Newton} \citep{2014A&A...563A...9S} and {\it Suzaku} \citep{2013PASJ...65...42H}. These studies find that the X-ray emission arises from a relatively hot ($kT\approx0.50-0.70$~keV), NEI plasma with enhanced abundances of Ne, Mg, and Si in several regions. Based on the derived ejecta masses, \citet{2013PASJ...65...42H} and  \citet{2014A&A...563A...9S} suggested that this remnant resulted from a $>$20M$_{\odot}$ progenitor.  \citet{2013PASJ...65...42H} also found hard X-ray emission spatially coincident with the molecular clouds and  the two OH masers \citep{koo05, 2013ApJ...771...91B,2014A&A...563A...9S}.

{\bf CTB~109 (G109$-$1.0)}--CTB~109, first detected in X-rays by {\it Einstein}  \citep{gregory80}, is the host of the anomalous X-ray pulsar (AXP) 1E 2259$+$586 \citep{gregory81}. Radio \citep{hughes81} and X-ray \citep{sasaki04} observations have revealed that CTB~109 has a semi-circular morphology, likely a result of interactions with a giant molecular cloud (GMC) complex on its entire western side \citep{sasaki13}. CO extends toward the east from the GMC complex and along the northern side of the SNR, and significant gamma-ray emission is detected throughout the SNR \citep{castro12}.

Another distinct feature of this remnant is the ``Lobe,'' a region of bright ejecta emission resulting from interactions with dense ISM. \cite{sasaki13} performed spectral analysis on the Lobe and surrounding regions using {\it Chandra} observations. They analyzed 39 regions in the northeast of the remnant, fitting them with both a single- and two-component NEI model. The two-component model represents the emission from both the shocked ISM and the ejecta. They found that the temperature of the ISM component was $kT\approx 0.1-0.3$~keV across all regions, and the temperature of the ejecta component was higher at $kT\approx 0.4-0.9$~keV. The ionization timescale of the ejecta component was $n_{\rm e}t \approx10^{11-12}$ s cm$^{-3}$ and they found absorbing columns of $N_{\rm H}\approx(0.5-1.0) \times 10^{22}$ cm$^{-2}$ in and west of the Lobe. Larger absorbing columns of $N_{\rm H}\approx(1.0-1.5)\times 10^{22}$ cm$^{-2}$ were found in the north where there is evidence of interaction with CO clouds \citep{sasaki06}. The spectral fits of \cite{sasaki13} revealed that the emission in CTB~109 is dominated by ISM below 0.6 keV and by ejecta above 1 keV. 

{\bf Cas~A (G111.7$-$2.1)}--Cas~A is one of the most studied remnants in the MW because it is the youngest CC SNR known in the galaxy with an age of only $\approx$340 years \citep{thorstensen01}. It was the target of {\it Chandra's} first light observation \citep{hughes00}, and since then many megaseonds of {\it Chandra} time have been dedicated to studying it. Prominent features of Cas~A are its distinct ejecta knots, which span from the center to and past the forward shock \citep{hughes00,hwang03}. Optical observations have shown that Cas~A is an O-rich SNR (e.g., \citealt{chevalier78}). Studies of light echoes from the explosion have revealed that Cas~A was produced by an asymmetric Type IIb supernova with variations in ejecta velocities of $\approx$ 4000 km s$^{-1}$ \citep{rest11}.

\cite{hwang12} and \cite{rutherford13} performed spectral analysis of several regions in Cas~A using {\it Chandra} ACIS observations, and modeled the data as an absorbed, NEI plasma. They found temperatures ranging from $kT\approx1-3$~keV and an average column density of $N_{\rm H}\approx1.3 \times 10^{22}$ cm$^{-2}$. The calculated abundances of Si, S, Ar, and Ca were 3--7 times the solar values, indicating that the emission from Cas~A is dominated by ejecta emission.

We note that the reverse shock in Cas~A has not yet propagated through the whole remnant \citep{gotthelf01}. However, \cite{hwang12} estimate that the unshocked ejecta only comprise $\approx$10\% of the total ejecta mass, so the X-ray emitting ejecta reflect the majority of the metals synthesized in the explosion.

{\bf CTA~1 (G119.5$+$10.2)}--CTA~1 was first detected in X-rays by {\it ROSAT} \citep{seward95} and has a center-filled X-ray morphology. It is a large SNR with a radio shell diameter of $\approx 1.8^\circ$ \citep{seward95} and a derived age of 1.3$\times 10^4$ yr from a Sedov solution \citep{slane04}. Non-thermal X-rays in CTA~1 reveal the presence of a PWN with a central radio pulsar RX~J0007.0$+$7303 that exhibits jet-like structure \citep{slane97,halpern04}. \cite{slane04} initially modeled the {\it ASCA} spectra of CTA~1 as an absorbed power-law and discovered that a thermal CIE component was required to fit the data below $\approx$1 keV. They found a column density of $N_{\rm H}\approx2.8 \times 10^{21}$ cm$^{-2}$, a spectral index of $\Gamma \approx2.3$, and a plasma temperature of $kT\approx 0.27$~keV. The non-thermal component arises from the synchrotron nebula that extends from the pulsar to the SNR shell. 

To date, the abundances of the thermal plasma have not yet been constrained, and thus it is unknown whether the thermal X-rays arise from the ejecta or the CSM/ISM. However, given the mature age of the SNR, it is likely that the X-ray emission is ISM-dominated.

{\bf IC~443 (G189.1$+$3.0)}--IC443 is another well studied mixed-morphology SNR. This source is coincident with both a giant molecular cloud complex \citep{1977A&A....54..889C} and an OB1 association \citep{1978ApJS...38..309H}. It is also among the brightest gamma-ray SNRs in the MW \citep[e.g.,][]{1996AAS...189.5507E, 1995A&A...293L..17S, 2009ApJ...698L.133A, 2013Sci...339..807A}. In addition to a number of radio studies \citep{1986A&A...164..193B, 2005ApJ...620..758S}, this remnant has been extensively followed up in X-rays \citep[e.g., ][]{1988ApJ...335..215P, 2002ApJ...572..897K, 2006ApJ...649..258T, 2008A&A...485..777T, 2009A&A...498..139B, 2009ApJ...705L...6Y, 2014ApJ...784...74O}. Originally, \citet{2008A&A...485..777T} suggested that the X-ray emission of IC~443 could be well described by a two-component plasma, one low temperature ($kT\approx0.30-0.50$~keV), NEI component associated with shocked ISM and a hot, ($kT\approx1.10-2.00$ keV) ejecta component in CIE with supersolar abundances of Mg, Si and S. However, analysis of {\it ASCA} and {\it Suzaku} observations revealed that an additional low temperature ($kT\approx 0.60$~keV), overionized plasma component is necessary to account for the line ratios and the radiative recombination continua features in the spectra \citep{2002ApJ...572..897K,2009ApJ...705L...6Y,2014ApJ...784...74O}.

{\bf Puppis~A (G260.4$-$3.4)}--Puppis~A was first detected in X-rays with {\it Einstein} observations \citep{petre82} and was subsequently observed with {\it ROSAT} \citep{aschenbach93}, {\it Chandra} \citep{hwang05}, {\it XMM-Newton} \citep{hui06,katsuda10}, and  {\it Suzaku} \citep{hwang08}. The X-ray emission in Puppis~A is thermal in nature \citep{luna16}, with evidence of knotty ejecta in the east and to the north of Puppis~A \citep{katsuda08,katsuda10,dubner13}. \cite{hwang08} performed spectral analysis using {\it Suzaku} observations, modeling various regions in the east of the remnant. They used an absorbed NEI model and found a temperature of $kT\approx0.6-0.8$~keV and a column density of $N_{\rm H}\approx0.3\times 10^{22}$ cm$^{-2}$. The northern regions showed slightly super-solar Si abundances, but most of the remnant is best-fit with sub-solar abundances, consistent with follow-up work demonstrating that the emission in Puppis A is dominated by CSM/ISM \citep{dubner13,luna16}. The emission on its eastern side is likely enhanced from interactions with a molecular cloud \citep{arendt10,dubner13}. In addition, there is evidence of dust absorption in the southwest from IR \citep{arendt10} and X-ray \citep{dubner13} observations, which may be the cause for the fainter X-ray emission in that region.

{\bf MSH 11$-$62 (G291.0$-$0.1)}--First observed with {\it Einstein} \citep{wilson86}, MSH~11$-$62 is a centrally-bright SNR with a young pulsar powering a PWN. The distance estimate to MSH~11$-$62 is uncertain, but thought to be near 5~kpc \citep{slane12}. In front of the SNR is the open cluster Tr 18 at a distance of $\approx1.5$~kpc \citep{vazquez90}, which can be seen as the many point sources in Figure~\ref{fig:snrs}. X-ray observations with {\it ASCA} \citep{harrus98}, {\it XMM-Newton}, and {\it Chandra} \citep{slane12} revealed that the spectrum of the remnant and pulsar can be best described using absorbed thermal and nonthermal components. \cite{slane12} fitted the spectra using a NEI plasma and a power-law component. They fixed the column density to the best-fit value for the PWN ($N_{\rm H}\approx6.7\times10^{21}$ cm$^{-2}$), and found an ionization timescale of $n_et\approx2.9\times 10^{10}$ s cm$^{-3}$ and $kT\approx2.8$ keV. There is evidence for super-solar abundances of Ne, Mg, and Si throughout the remnant, indicating that the thermal emission reflects the distribution of ejecta.

{\bf G292.0$+$1.8}--G292.0$+$1.8 is an ``oxygen-rich'' SNR based on optical emission from fast ejecta knots rich in oxygen \citep{goss79, murdin79}. It has an elliptical X-ray morphology with a central equatorial belt running east to west \citep{park02,lee09,ghavamian16}. Several X-ray studies using {\it Chandra} \citep{park04, bhalerao15} have shown that G292.0$+$1.8 has emission from both metal-rich ejecta and shocked CSM, with CSM mainly located in the equatorial belt. \cite{bhalerao15} performed spectral modeling of 33 regions. They modeled the spectra as an absorbed NEI plasma plus a power-law component for regions near the central pulsar. They allowed O, Ne, Mg, Si, S, and Fe abundances to vary and found that most regions in the remnant have super-solar abundances ($\gs$ several times solar), indicating an ejecta origin of the X-ray remission. However, most of the emission from the equatorial belt has sub-solar abundances and is likely CSM material. 

{\bf PKS 1209$-$51 (G296.5$+$10.0)}--First detected in X-rays with {\it HEAO-1} \citep{tuohy79}, the SNR PKS~1209$-$52 is a large (81$\arcmin$-diameter) SNR with a radio-silent central NS \citep{zavlin98}. The NS, 1E~1207.4$-$5209, was first discovered with {\it Einstein} \citep{helfand84}. \cite{kellett87} performed spectral analysis on {\it EXOSAT} data and found evidence for thermal emission. They used an absorbed single-temperature CIE plasma model, and found a column density of $N_{\rm H}\approx1.4\times 10^{21}$ cm$^{-2}$ and an average temperature of $kT\approx0.15$~keV. However, due to its large and diffuse nature, {\it Chandra}, {\it XMM-Newton}, and {\it Suzaku} have not mapped the entirety of this SNR; most observations have focused on the vicinity of the CCO. Thus, the abundances of the thermal plasma have not been measured previously and the thermal emission could arise from ejecta or CSM/ISM. 

PKS 1209$-$51 does show signs of interaction. In particular, \cite{araya13} analyzed {\it Fermi} Gamma-ray Space Telescope observations of this SNR, and it was detected with $\approx$5-$\sigma$ significance above 500~MeV. Many GeV-bright SNRs are known to be interacting with molecular clouds (e.g., \citealt{hewitt09,castro13}), so the {\it Fermi} detection of PKS~1209$-$51 may suggest that the remnant is interacting. It is unclear whether or how this remnant is interacting with CSM or ISM, so the emission could reflect the ejecta distribution or the local environment.

{\bf MSH 15$-$56 (G326.3$-$1.8)}--First detected in radio, \citep{mills61}, MSH~15$-$56 is a composite SNR. Radio observations revealed a SNR shell and a luminous PWN \citep{dickel00}.  It was also detected in X-rays by {\it ROSAT} \citep{kassim93}, {\it ASCA} \citep{plucinsky98}, {\it XMM-Newton} and {\it Chandra} \citep{temim13}. It has clumpy X-ray emission, with enhancement near the PWN. \cite{temim13} performed spectral analysis of eight regions in the southwest area of MSH 15$-$56 using {\it Chandra} and {\it XMM-Newton} observations, focusing on the pulsar and the surrounding regions. An absorbed, two-component model with a power-law and a NEI plasma was needed to accurately fit the spectra. They found a best-fit column density of $N_{\rm H}\approx5.1\times 10^{21}$ cm$^{-3}$, a temperature of $kT\approx0.5-0.6$~keV, and an ionization timescale of $n_{\rm e}t\approx2\times 10^{11}$ s cm$^{-3}$ for most regions. The northern regions, the largest and most central to the SNR, showed enhanced abundances of Si and S (3--4$\times$ solar values: \citealt{temim13}). Regions near the pulsar showed abundances closer to solar values or were unconstrained by the fit. 

{\bf RCW~103 (G332.4$-$0.4)}--RCW~103 is a shell-type remnant with an estimated age of $\approx 2000$ years \citep{carter97}. It is host to an unusual magnetar with a 6.67-hour periodicity thought to be generated by a fall-back accretion disk \citep{tong16,ho17,dai16} or a binary \citep{deluca06}. \cite{oliva90,oliva99} performed IR spectroscopy of RCW~103 and found emission from molecular and ionized gas in the south. Further studies in the IR and radio revealed that this remnant is interacting with a molecular cloud in the southeast \citep{pin11}. 

In the X-rays, \cite{lopez11} and \cite{kari15} performed spectral analysis of RCW~103 using {\it Chandra} observations. \cite{lopez11} analyzed 31 regions and modeled the spectra as an absorbed NEI plasma. They found an average column density of $N_{\rm H}\approx 6\times 10^{21}$ cm$^{-2}$, a temperature of $kT\approx 0.53$ keV, and an ionization timescale of $n_{\rm e}t\approx7 \times 10^{11}$ s cm$^{-3}$. In nearly all regions examined, the abundances of Mg and Si were found to be statistically above solar values, except in the southeast. \cite{kari15} also modeled the {\it Chandra} spectra with one NEI or with two thermal (one NEI and one CIE) components. They found that the single-component fits suggested the CSM dominates the X-ray emission (based on the subsolar best-fit abundances). However, in the two-component fits, they found that roughly half of the 27 regions analyzed have ejecta emission (with supersolar abundances of Ne, Mg, Si, S, and/or Fe). \cite{kari15} noted that it is challenging to disentangle the CSM and ejecta emission in RCW~103 as they are interspersed throughout the SNR.

\section{Methods} \label{sec:methods}
We use the power-ratio method (PRM) to analyze the asymmetries of X-ray emission in SNRs. This method was employed to characterize the X-ray morphology of galaxy clusters \citep{buote95,buote96,jeltema05} and was adapted by \cite{lopez09a} for use on SNRs. Using the PRM, we calculate the multipole moments of extended emission to probe asymmetries. Power-ratios, the nth-order power moment divided by the zero-th power moment, are used to directly compare sources of different total fluxes. For a more detailed/mathematical description of this method and its application to SNRs, see \cite{lopez09a}. 

\subsection{Center-of-Emission Analysis}
In our first series of calculations, we adopt the center-of-mass of the emission (hereafter center-of-emission) as the origin of the multipole expansion. In this case, the dipole power-ratio, P$_1$/P$_0$, approaches zero while the higher-order moments give details about successively smaller-scale asymmetries. P$_2$/P$_0$ is the quadrupole power-ratio and quantifies ellipticity/elongation of an extended source. P$_3$/P$_0$ is the octupole power-ratio and is a measure of mirror asymmetry. Figure~\ref{fig:snrs} shows the center-of-emission (blue crosses) for each SNR.
 
Uncertainties in the power-ratios are estimated via the Monte Carlo analysis described in \cite{lopez09a}. The program \emph{AdaptiveBin} \citep{sanders01} is used to bin the SNR into sections of equal total photon counts. Then, the counts in each bin are replaced by a number taken randomly from a Poisson distribution with the mean equaling the original number of counts. This process is repeated 100 times to create 100 mock images of each source. We then measure the power-ratios of these 100 images, and take the average as the true power-ratio. We adopt the sixteenth-highest and -lowest values as the confidence limits, chosen to match the 1-$\sigma$ range of a Gaussian distribution.

\subsection{Explosion Site Analysis}
For the sources with reliable measures of the explosion sites (Cas~A, G292.9$+$1.8, CTB~109, Puppis~A, PKS~1209, and RCW~103), we also compute the power-ratios by adopting the explosion sites as the origin of the multipole expansion. In this case, the dipole power-ratio (P$_1$/P$_0$) measures the displacement between the center-of-emission and the explosion site.

The explosion sites are derived by back-evolving the proper motion of the NS or SNR filaments assuming ages given in the literature (listed in Table~\ref{table:SNRtable}). The yellow ellipses in Figure~\ref{fig:snrs} show the explosion site locations with sizes reflecting the 1-$\sigma$ confidence region using the back-evolved NS velocities or filaments. For some sources (e.g., G11.2--0.3, G189.1$+$0.3), this method results in explosion sites that are unphysical, such as at the edge of the SNR shell, even though statistical uncertainties in the NS proper motion are small. Systematic uncertainties may be substantially larger, as in e.g., G11.2$-$0.3, where the ellipse is significantly offset from the center-of-emission and the geometric center. For these sources, we do not use the proper motion velocity. Other sources do not have precise enough NS or filament proper motions to determine reliable explosion sites. For example, G34.7$-$0.4 has large errors that provide almost no constraint on the explosion site, and both Kes~73 and MSH~11$-$62 have velocity errors of $< 100$ km s$^{-1}$. Consequently, for these sources, we do not perform the power-ratio analysis using the explosion site as the origin.

  \begin{figure*}
   \includegraphics[width=0.49\textwidth]{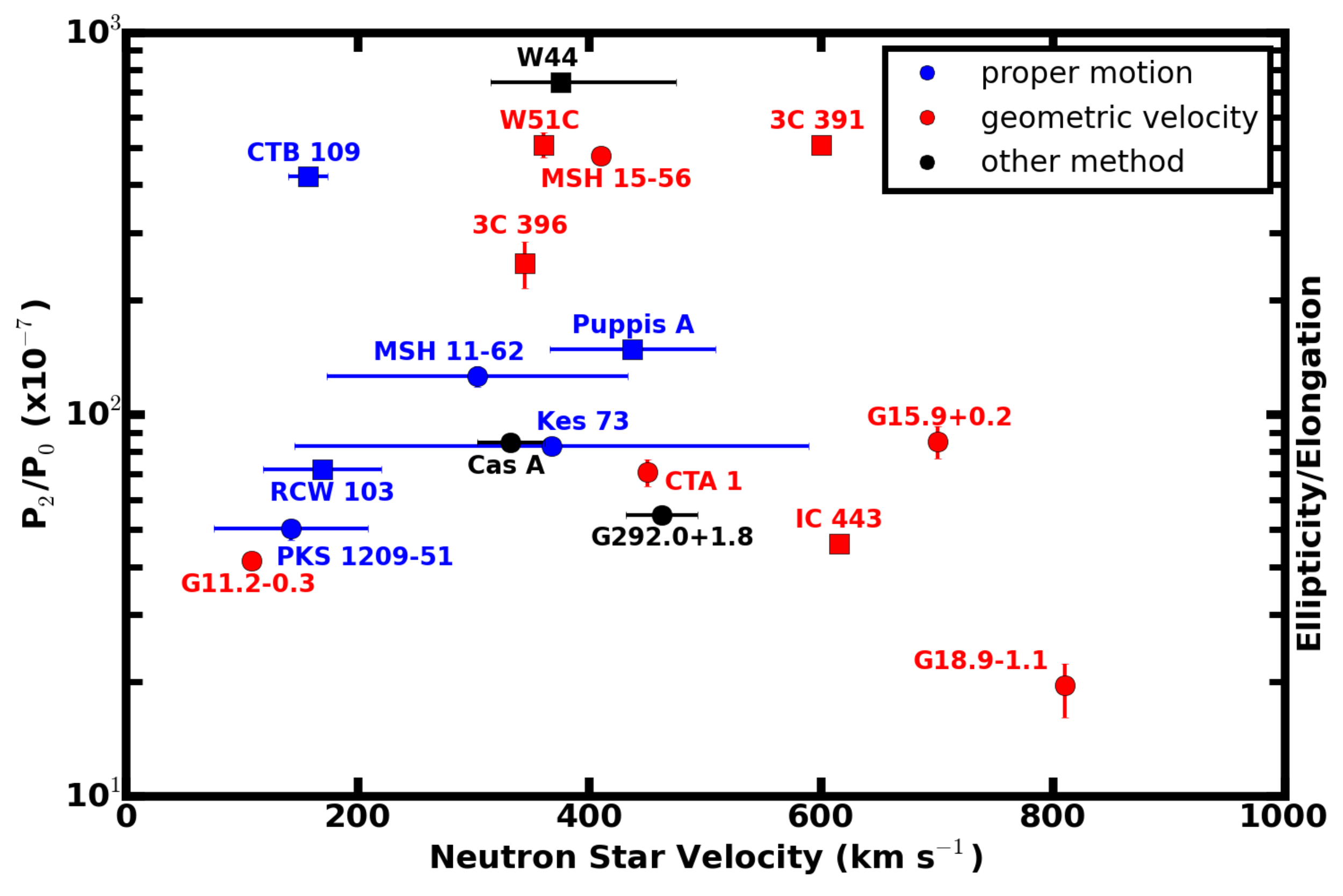}
 \includegraphics[width=0.49\textwidth]{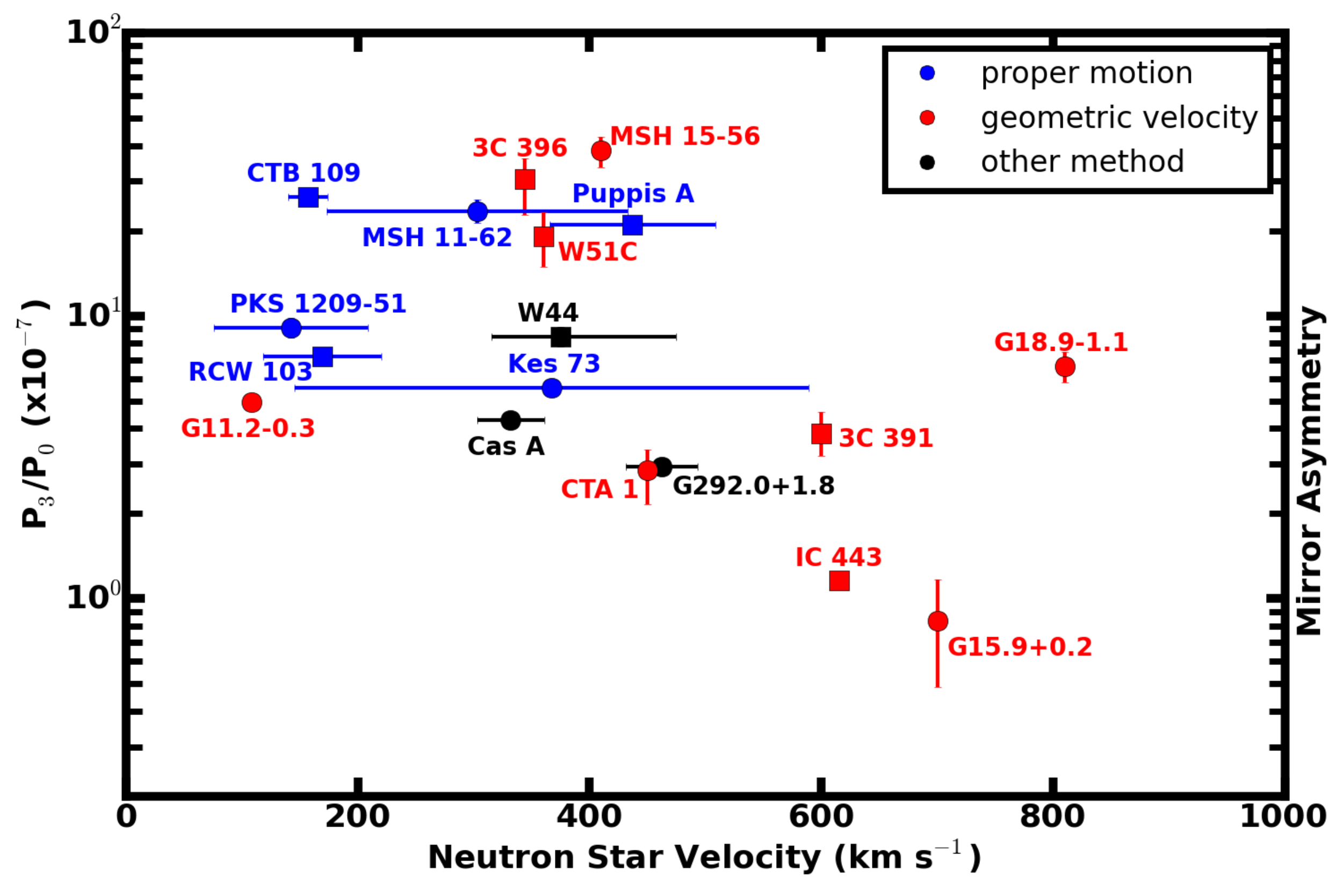}
  \caption{The quadrupole (left) and octupole (right) power-ratios vs. neutron star velocities for the sample of SNRs. Blue points are velocities measured via proper motions while red points represent velocities measured via the geometric method. Cas A and G292.0$+$1.8's NS velocities are determined by back-evolved filament motion to find the explosion site, while W44's NS velocity is determined by multiple methods described in Section~\ref{subsec:nsvel}. Circles indicate there is no evidence of SNR interaction with CSM/ISM, and squares indicate clear evidence of interaction. The error bars associated with the velocity measurement for the red points only reflect the uncertainty in the geometric center. They do not account for age and distance uncertainties or the difference between geometric center and explosion site, all of which lead to an error likely $\gs 2$ times the reported velocity. }
  \label{fig:com}
 \end{figure*}
 
\subsection{Neutron Star Velocities} \label{subsec:nsvel}

Generally, two approaches are utilized to estimate NS velocities in literature: 1) a proper motion measurement using multi-epoch images (e.g., \citealt{auchettl15}); 2) approximation of the velocity, given the NS's spatial offset from the geometric center and the age and distance of the SNR (e.g., \citealt{kaspi01}). The latter approach requires a single observation with a NS detection; thus it is the most common way to estimate velocities in the literature. However, this method can lead to significant errors as it assumes that the NS originated at the geometric center of the SNR. This assumption is not necessarily valid as interactions with an inhomogenous medium can distort the shape of SNRs so that the geometric center is not the actual explosion site. Furthermore, an assumption of age is necessary to convert geometric offsets to NS velocities, but SNR dynamical ages are often uncertain by a factor of $\gs 2$ (e.g., G350.1$-$0.3 has a calculated age of 600--1200 years: \citealt{lov11}) and have correlated errors with SNR distance estimates. In addition, growing evidence shows that spin--down ages of NSs might not represent the true ages of the sources \citep{popov12,nakano15,rogers17}. 

Proper motion measurements yield more accurate velocities but require multiple epochs of high spatial resolution images over a long time baseline. The average transverse velocity of isolated radio pulsars is close to 250 km s$^{-1}$ \citep{hobbs05}, which corresponds to a motion of only $\approx$0.2\arcsec\ over a 10-year baseline for a Milky Way remnant at a distance of 3 kpc. Thus, sub-arcsecond spatial resolution is crucial for measuring NS proper motions, and {\it Chandra} is the only current X-ray facility with that capability.

We searched the literature for reported NS velocities associated with Galactic SNRs, preferably using proper motions, but including those estimated using the geometric offset method. NS proper motions have been accurately measured in four SNRs that meet our selection criteria (CTB 109, IC~443, Puppis A, and PKS~1209$-$51) using {\it Chandra} or near-infrared imaging. These values and associated references are listed in Table~\ref{table:SNRtable}. 
 
In addition, three SNRs (Cas~A, G292.0$+$1.8, and W44) in our sample have NS velocities measured in the literature using methods other than the geometric offset or NS proper motion, and we adopt these values in our analysis. The explosion sites of Cas~A and G292.0$+$1.8 were inferred from the motion of filaments detected through extensive monitoring of these SNRs \citep{thorstensen01,fesen06,winkler09}. Given the inferred distances and ages to these sources, we calculate the transverse velocities of their associated NSs using the inferred explosion sites and the current position of the NSs.

The velocity of the NS B1953$+$01 in the SNR W44 was estimated using multiple methods resulting in velocities of 315--470 km s$^{-1}$ with a median value of 375 km s$^{-1}$ \citep{frail96}. These methods include the geometric offset method, a relation between the NS's velocity and the SNR shock velocity, and balancing pressure between the PWN and the hot gas interior in W44 (see \citealt{frail96} for more details).

For all other targets, we searched for multi-epoch {\it Chandra} archival observations to derive our own transverse NS velocities. Five sources (G11.2$-$0.3, G15.9$+$0.2, Kes~73, MSH~11$-$62, and RCW~103) had multiple ACIS observations over a baseline of $\gs$10 years (see Table~\ref{table:obslog} for details). For each SNR, we used the two observations over the longest baseline in which the NS was detected to measure the NS proper motion. The observations used are listed in Table~\ref{table:obslog}, denoted by the rows with ``NS'' as the purpose. 

The astrometric accuracy of {\it Chandra} ACIS\footnote{See http://cxc.harvard.edu/cal/ASPECT/celmon/} is $\sim$0.6\arcsec, which is of the same order of magnitude as the expected NS motion over the $\sim$10 year baselines. To improve this accuracy, we performed an astrometric correction using the positions of point sources across two epochs of {\it Chandra} observations (a similar method as in e.g., Section 2 of \citealt{becker12}). We identified point sources detected in all epochs of observations and separated them into two categories: sources associated with optical stars from the UCAC4 catalog \citep{zach13} and sources likely to be active galactic nuclei (AGN) based on their high X-ray hardness ratios. We derived the true positions of the stars at both epochs, accounting for their proper motions, and assumed that AGN, as background sources, do not move. We calculated an error-weighted average of the difference in identified point sources' known positions and their {\it Chandra} positions (as measured using the {\sc ciao} command {\it wavdetect}) and used this value to correct the position of the NSs at each epoch. These corrected positions are used to calculate the NS's motion over the two epochs. The velocities, along with their errors and directions, are listed in Table~\ref{table:SNRtable}.

This method produces uncertainties of tens to hundreds of km s$^{-1}$ and is likely more representative of the true velocities than those derived from geometric offsets. If the resulting proper motions were consistent with zero km s$^{-1}$ at the 1-$\sigma$ level and/or the velocity is significantly above 1000 km/s at the 1-$\sigma$ level (as was the case for G11.2$-$0.3 and G15.9$+$0.2), then we adopted the velocity given by the geometric offset method. For the remaining remnants that did not have multi-epoch {\it Chandra} observations, we estimated the NS velocities using their offsets from the geometric center of the SNR and assume the SNRs' dynamical ages are the NSs' ages (listed in Table~\ref{table:SNRtable}).

\subsection{Checks}

 \begin{figure*} 
  \includegraphics[width=0.32\textwidth]{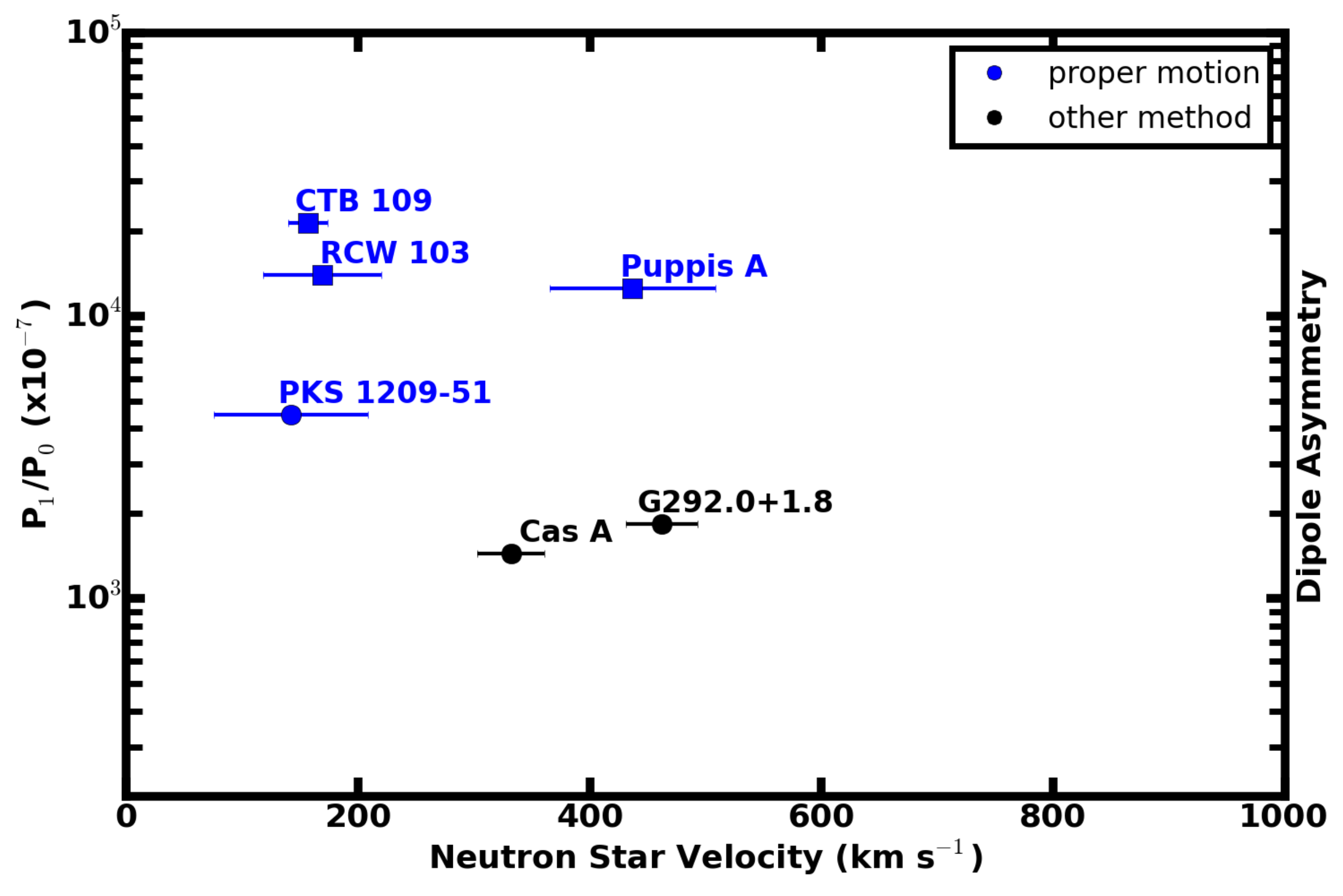}
 \includegraphics[width=0.32\textwidth]{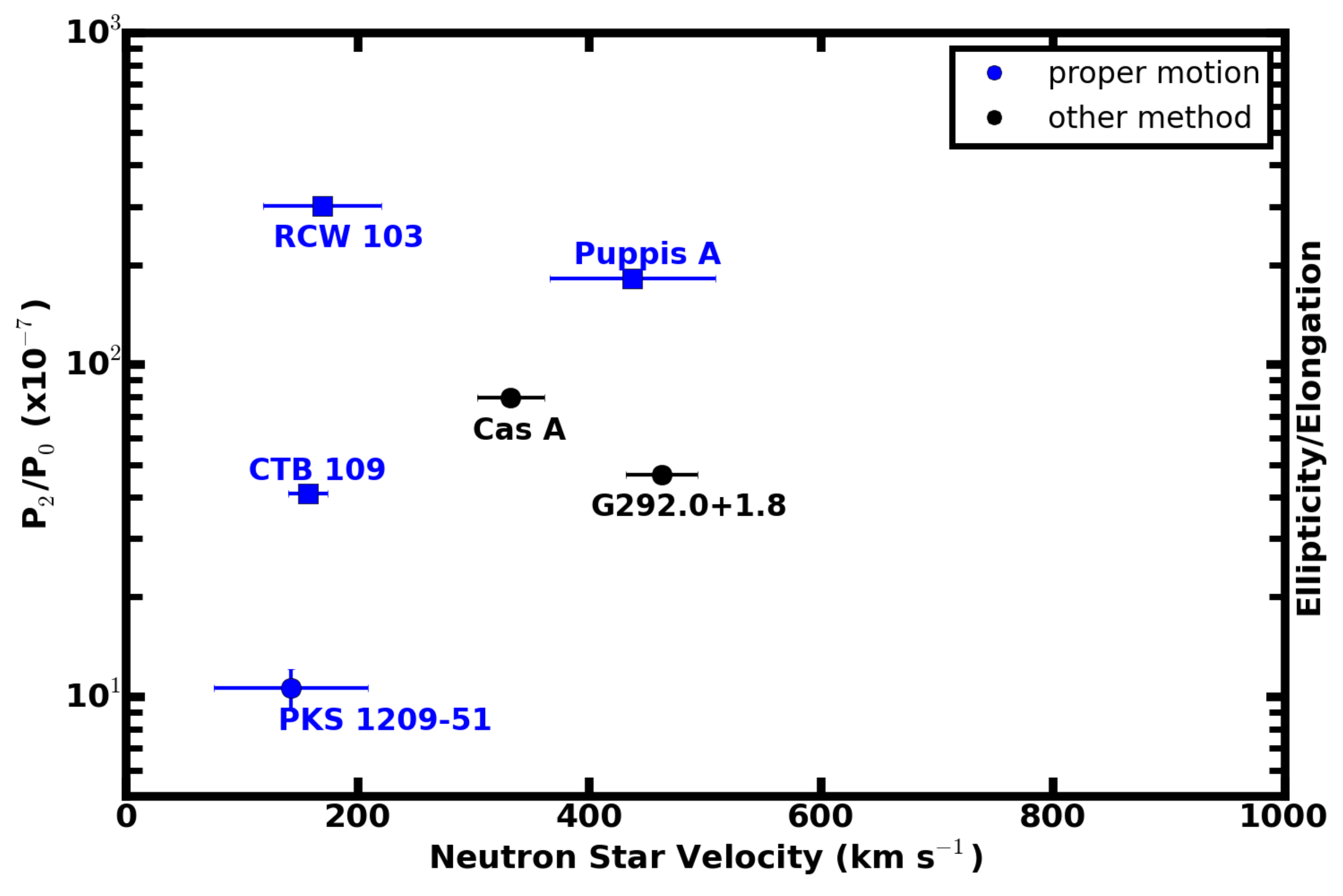}
 \includegraphics[width=0.32\textwidth]{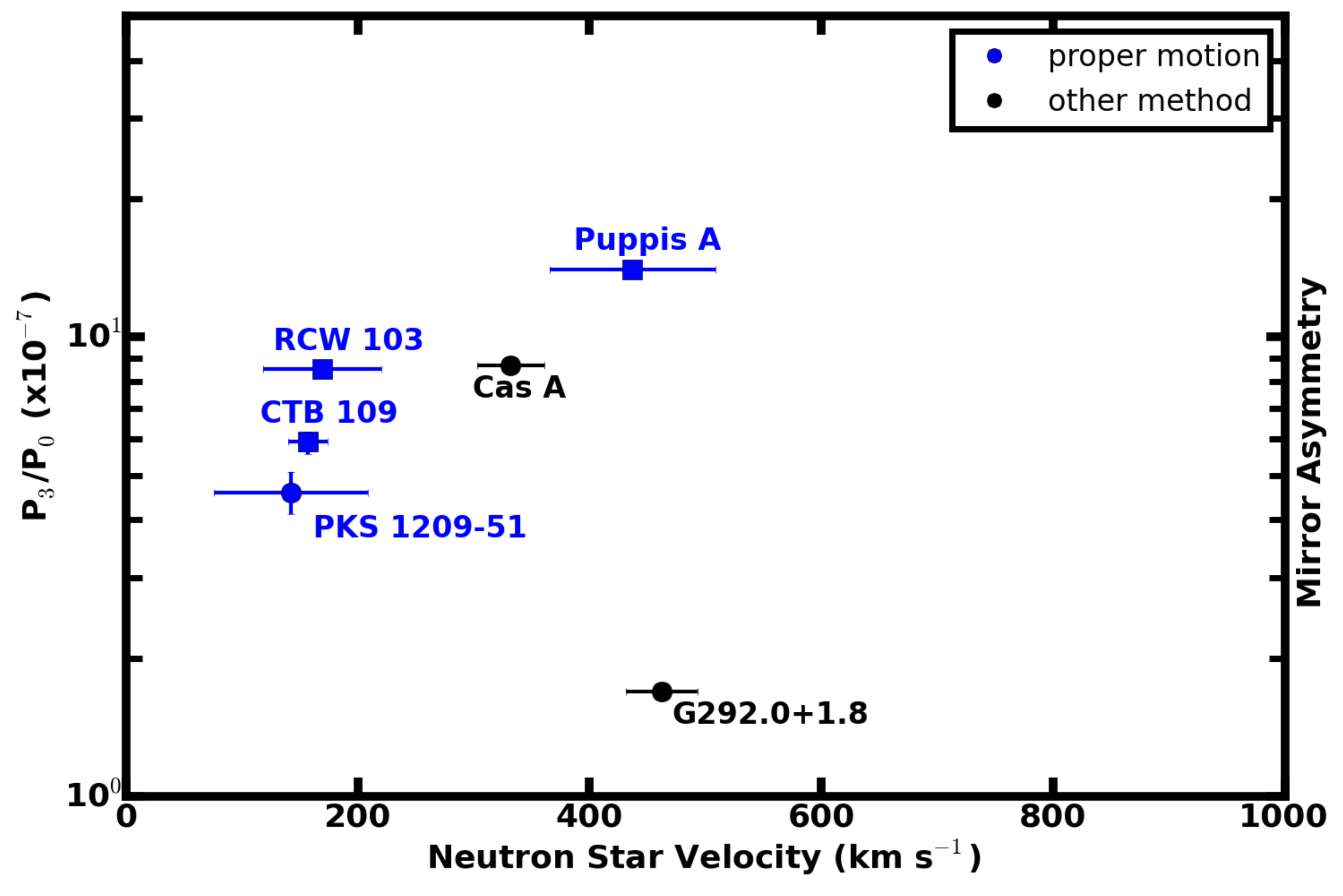}
\caption{The dipole (left), quadrupole (middle), and octupole (right) power-ratios vs. neutron star velocities for the sample of SNRs, using the explosion site as the origin for analysis. Blue points are for NS proper motion determined velocities and explosion sites while black points are for filament proper motion determined velocities and explosion sites (Cas~A and G292.0$+$1.8). Circles indicate there is no evidence of SNR interaction with CSM/ISM, and squares indicate clear evidence of interaction. 
}
  \label{fig:origin}
 \end{figure*}
To verify the robustness of the PRM, we compared the derived power-ratios for individual sources observed with both {\it Chandra} and {\it ROSAT}. These values were consistent across our sample. We also explored whether the power-ratios correlated with physical size or age of the SNRs, and we found no statistically significant trends between any of these parameters.

Furthermore, we investigated whether the estimated NS velocities depended on the time baseline between the multi-epoch observations. For example, the reported velocity of the NS in Puppis~A has decreased as observations with longer baselines were used, going from 1122 to 672 km s$^{-1}$ with 5.5- and 10.5-year baselines, respectively \citep{hui06,becker12}. In our sample, we note that no NS proper motions with $\gs$10 year baselines had velocities over 1000 km s$^{-1}$. Thus, long baselines ($\gs$10 years) reduce systematic uncertainties and are crucial to derive reliable NS proper motions.
 
\section{Results} \label{sec:results}

Figure~\ref{fig:com} shows the quadrupole power-ratio (left) and octupole power-ratio (right) versus the NS transverse velocities using the center-of-emission as the origin of the multipole expansion. We find no correlation between either $P_2/P_0$ or $P_3/P_0$ and NS velocity. This lack of a significant correlation holds for both the proper motion and geometric offset velocity estimates when taking into account the large errors of the geometric estimates. In addition, both non-interacting and interacting SNRs (presented as circles and squares in Figure~\ref{fig:com}, respectively) do not show a trend in the power-ratios versus NS velocity. However, we note that most SNRs with large $P_{2}/P_{0}$ and $P_{3}/P_{0}$ are known to be interacting.

When we adopted the explosion site as the origin of the multipole expansion, we also found no correlation between the power-ratios (dipole, quadrupole, or octupole) and the NS velocities (see Figure~\ref{fig:origin}). We note that the interacting SNRs tend to have larger dipole power-ratios than the non-interacting SNRs. However, the sample size of six SNRs in this analysis is small. 

In addition to examining the magnitude of the power-ratios and NS velocities, we also investigated their directions. Each power-ratio -- the dipole (P$_1$/P$_0$), quadrupole (P$_2$/P$_0$), and octupole (P$_3$/P$_0$) -- has an associated direction. The dipole power-ratio angle points toward the bulk of the X-ray emission. The quadrupole power-ratio angle represents the semi-major axis of the ellipse that best matches the SNR's X-ray emission. The octupole power-ratio angle points in the direction of the largest mirror asymmetry. 

   \begin{figure*}
      \includegraphics[width=\textwidth]{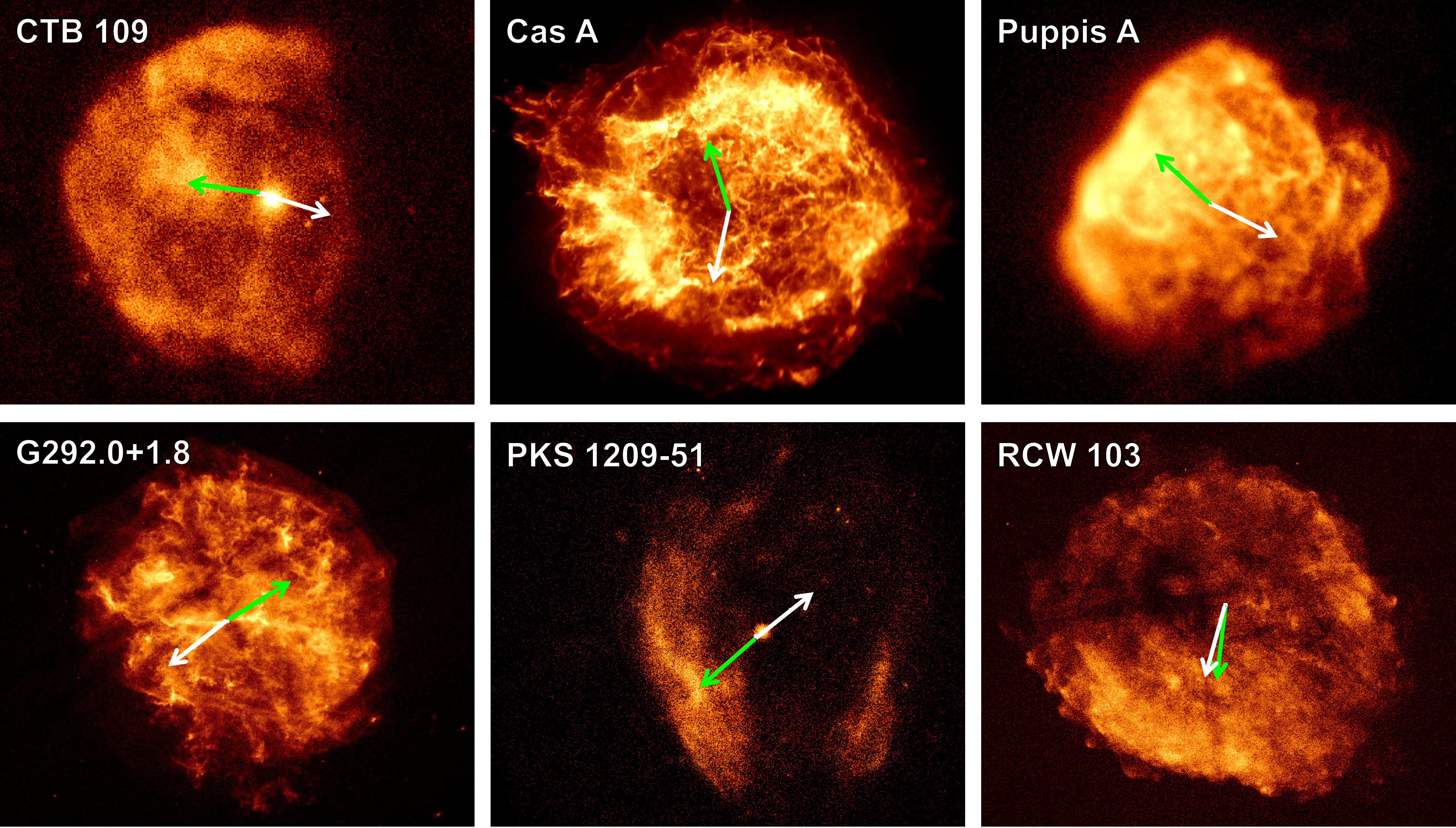}

 \caption{0.5--2.1 keV {\it Chandra} and {\it ROSAT} images of the six SNRs for which we have robust measures of their explosion sites. The green arrow points from the explosion site to the direction of the dipole moment. The white arrow points in the direction of NS motion. In Cas~A and G292.0$+$1.8, the dipole moment direction reflects the bulk of ejecta emission. In CTB~109, Puppis~A, and RCW~103, the dipole moment points towards enhanced emission due to interactions with CSM/ISM or a molecular cloud. It is unclear if the emission comes from ejecta or interactions in PKS~1209$-$51.}
  \label{fig:NS_dipole} 
 \end{figure*}
 
We compared the angle associated with each multipole term to the direction of the NS's motion. For this analysis, we only used the six SNRs that had accurate explosion sites determined by NS or filament proper motion measurements (CTB~109, Cas~A, Puppis~A, G292.0$+$1.8, PKS~1209$-$51, RCW~103). Figure~\ref{fig:NS_dipole} shows images of these SNRs with their dipole angle and NS direction of motion labeled. We do not perform this calculation using the center-of-emission as the origin for the power-ratio analysis because the NS does not originate from the center-of-emission. Thus the difference in angles has no physical significance.

\begin{figure*}
 \begin{center}
 \includegraphics[width=\columnwidth]{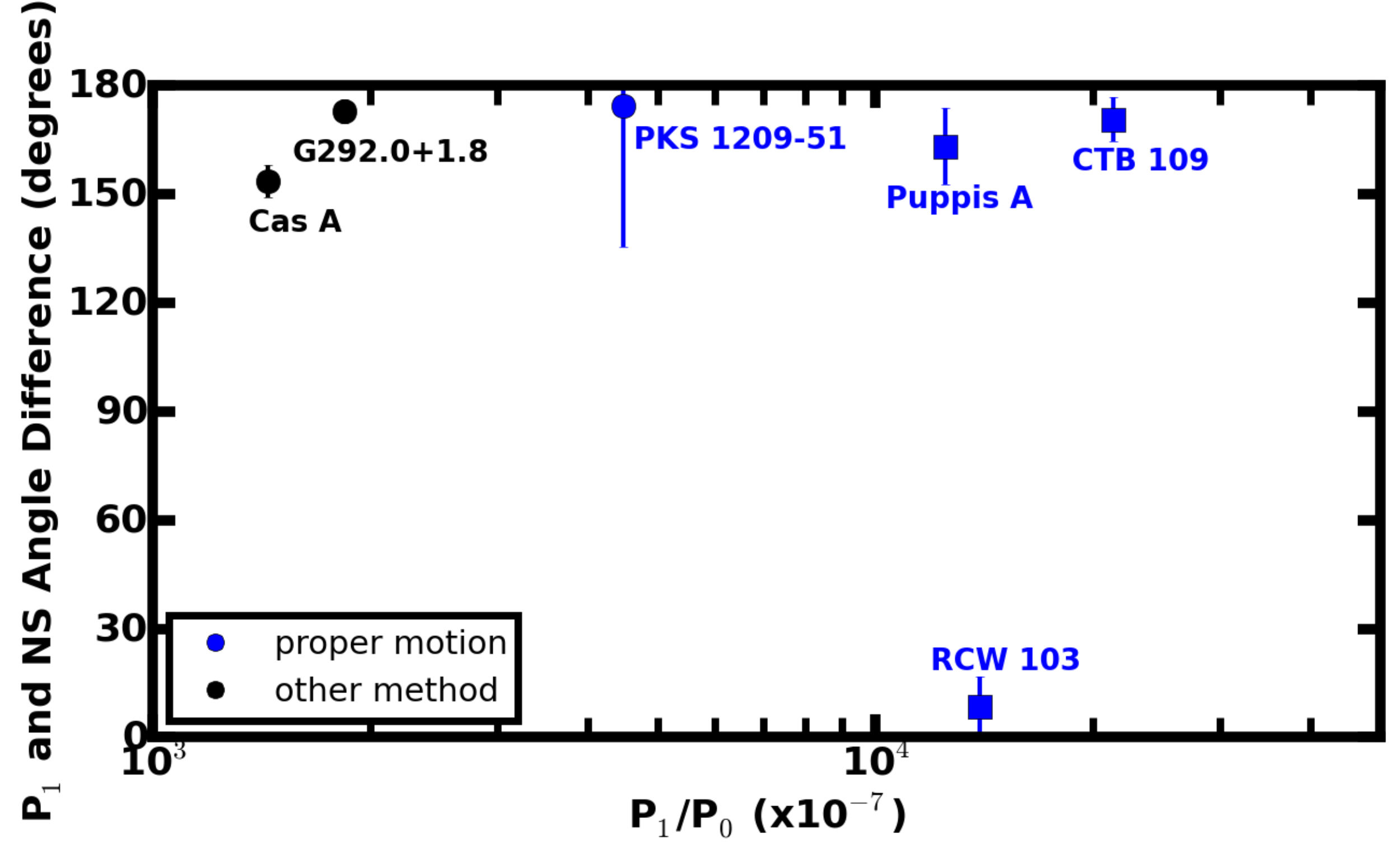}
 \end{center}
 \caption{The angle difference between the dipole angle and the direction of NS motion from the SNR explosion site as a function of the magnitude of the dipole power-ratio. The explosion sites for Cas~A and G292.0$+$1.8 are calculated using back-evolved filament motion, which is then taken as the NS birth site. The explosion sites for the rest are determined by back evolving the NS's proper motion.  Circles indicate there is no evidence of SNR interaction with CSM/ISM, and squares indicate clear evidence of interaction.}
  \label{fig:p1angle}
 \end{figure*}

\begin{figure*}
 \begin{center}
 \includegraphics[width=\columnwidth]{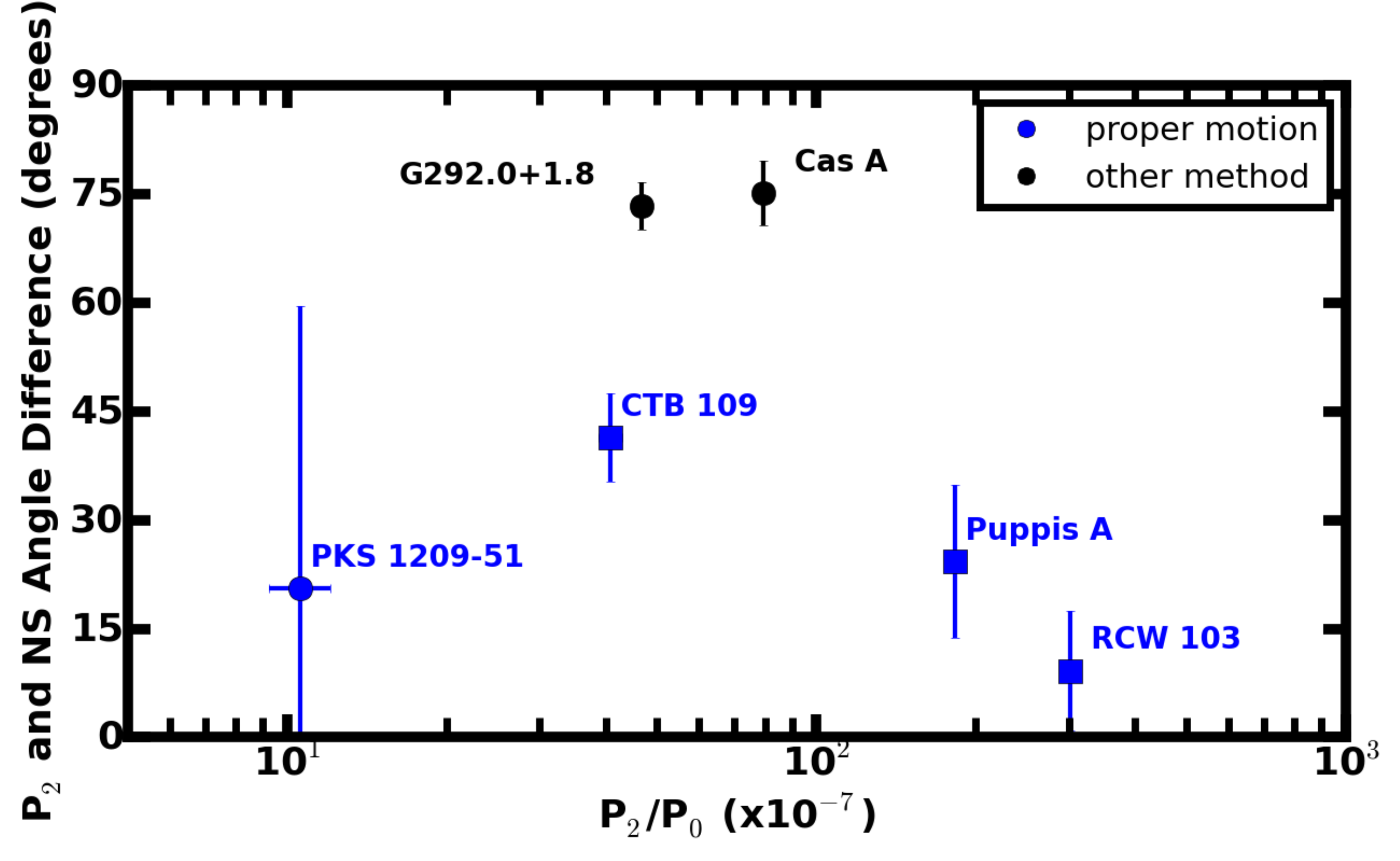}
  \includegraphics[width=\columnwidth]{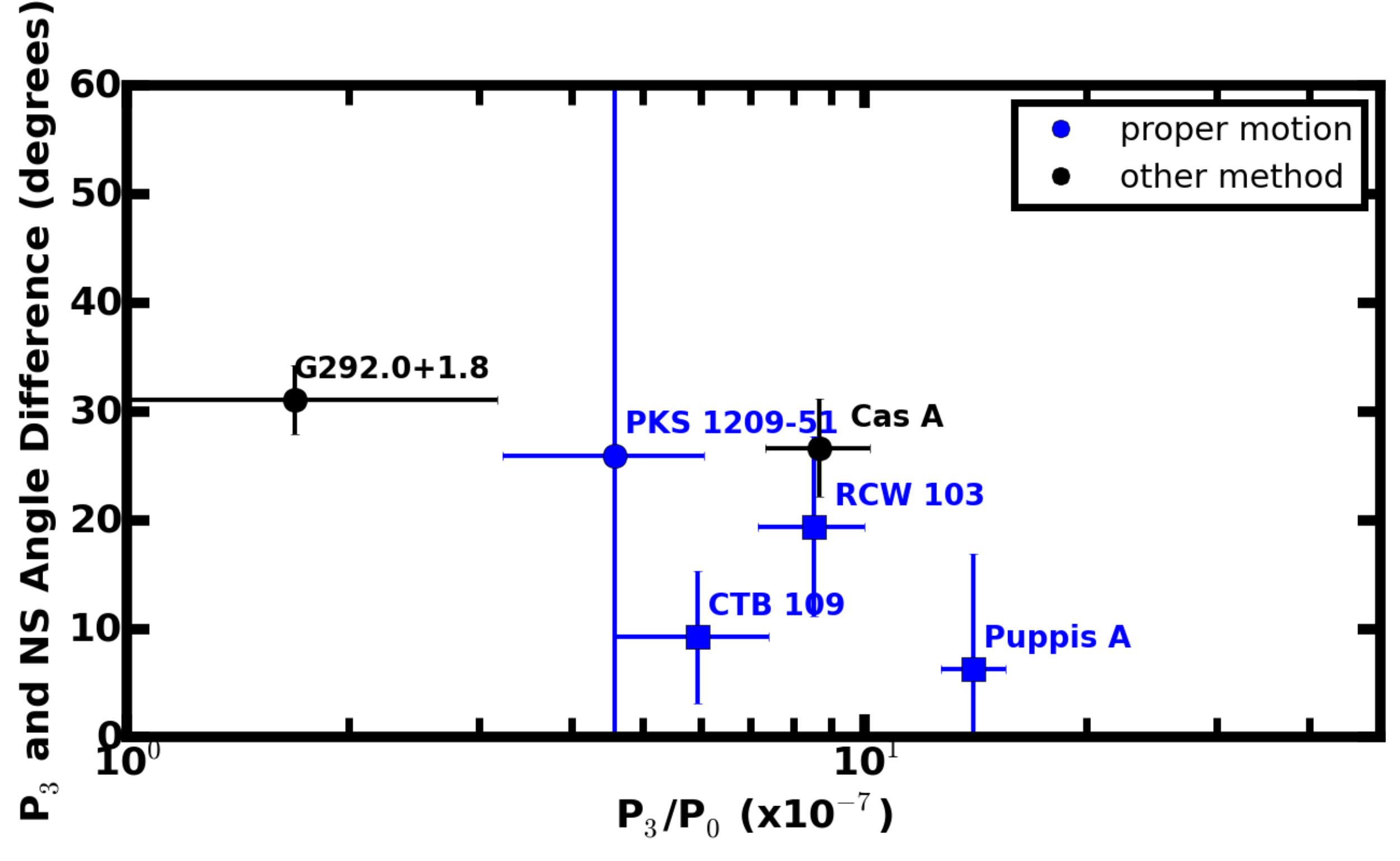}
 \end{center}
 \caption{The angle difference between the direction of the quadrupole (left) and the octupole (right) moments and the direction of neutron star movement from the SNR explosion site as a function of the magnitude of the power-ratios. The explosion sites for Cas~A and G292.0$+$1.8 are calculated using back-evolved filament motion, which is then taken as the NS birth site. The explosion sites for the rest are determined by back-evolving the NS's proper motion. Circles indicate there is no evidence of SNR interaction with CSM/ISM, and squares indicate clear evidence of interaction.}
  \label{fig:p2p3angle}
 \end{figure*}

Figures~\ref{fig:p1angle} and \ref{fig:p2p3angle} show the difference in the direction of the dipole, quadrupole, and octupole power-ratios and the direction of NS propagation, as a function of the respective power-ratios. Five of the six SNRs with well-constrained explosion sites (Cas~A, G292.0$+$1.8, CTB~109, Puppis~A, and PKS~1209$-$51) yielded dipole angle differences of $\sim$155--180$^\circ$, indicating that the NS is moving in the opposite direction as the majority of emission. However, in RCW~103, the NS is moving in the same direction as the dipole angle. We see no statistically significant trend for either the quadrupole or octupole power-ratio angle measurements. In the next paragraphs, we discuss each of the six sources of Figure~\ref{fig:p1angle} in turn, focusing on whether emission arises from ejecta or CSM/ISM. If the remnants' emission are ejecta dominated, then the data in Figure~\ref{fig:p1angle} reflect the direction of ejecta compared to the NSs' motion and thus can be compared to the predictions of SN explosion simulations. Otherwise, the emission may trace the CSM/ISM and be unrelated to the explosion and kick mechanism.

The dipole of Cas~A points to the north/northeast, while the NS is moving to the south/southwest, producing an angle difference of $\sim$155$^{\circ}$. This result is consistent with the finding of \cite{hwang12} who showed that the ejecta are recoiling in a direction that is $\sim$150$^{\circ}$ from the NS's motion. As the youngest CC SNR known in the Milky Way, Cas A's emission is ejecta-dominated \citep{hwang12,rutherford13}, and thus its morphology reflects the asymmetries of the explosion.  

G292.0$+$1.8 is another SNR with ejecta-dominated emission where the NS direction (moving to the southeast) is opposite to the bulk of emission toward the northwest. As discussed in Section~\ref{sec:csm}, G292.0$+$1.8 has shock-heated emission from CSM/ISM along the equatorial belt, but the dipole angle points to ejecta in the northwest. Thus, we conclude that the dipole angle reflects the bulk distribution of ejecta in G292.0$+$1.8.

PKS~1209$-$51 (G296.5$+$10.0) is the third SNR where the majority of the emission (to the southeast) is opposite to the NS's motion (to the northwest). Although the SNR is bright in GeV gamma-rays \citep{araya13}, no other indicators of molecular cloud interaction have been found. Thus, it is unclear whether inhomogeneous heating from CSM/ISM interactions has altered the SNR's X-ray morphology and affected the dipole angle direction. In addition, while the X-ray emission originates from a thermal plasma \citep{kellett87}, the abundances cannot be constrained with existing data, so the thermal emission could arise from ejecta or CSM/ISM. 

Puppis~A (G260.4$-$3.4) also has the majority of emission (to the northeast) opposite to the NS's motion (to the southwest). The dipole angle points to a region with super-solar abundances, indicating an ejecta origin for that emission \citep{dubner13}. However, it is likely enhanced due to an interaction with a molecular cloud there. Furthermore, as discussed in Section~\ref{sec:csm}, the integrated emission of the SNR is dominated by CSM/ISM overall. Thus, the X-ray morphology is influenced by the environment and may not reflect the distribution of the ejecta.

For CTB~109, the dipole direction is oriented toward the east while the NS is moving to the west. Spectral models by \cite{sasaki13} demonstrate that the SNR has significant ejecta emission in the 0.5$-$2.1 keV band. However, enhanced emission in the northeast results from interaction with ISM, and the west is dim because of a GMC complex there. Thus, the X-ray morphology may be reflective of its numerous interactions.

The NS in RCW~103 is moving in the same direction as the dipole angle (toward the southeast), contrary to the results from the other SNRs. The most likely reason for this result is that RCW~103 is interacting with a molecular cloud to the south \citep{kari15}, enhancing X-ray emission there. As \cite{kari15} discussed, the ejecta emission seems well-mixed with the CSM/ISM emission. There are no regions where ejecta dominates the emission relative to that from the CSM/ISM (like there are in CTB~109 and Puppis~A), so the molecular cloud interaction plays the largest role in governing the dipole direction. Thus, the $\sim$0$^\circ$ angle difference may not reflect the relationship between the ejecta distribution and NS motion. 

In addition, we note that the NS in RCW~103 has an unusual 6.67-hour periodicity, whose formation history is unknown (e.g., \citealt{deluca08b,tong16,ho17,dai16}). Thus, an unusual explosion may have also affected the relationship between the bulk of the ejecta and the NS kick.

\section{Discussion} \label{sec:disc}

We find no correlation between the NS velocities and the magnitude of the power-ratios in the 0.5--2.1 keV images. This result holds whether we adopt the center-of-emission (Figure~\ref{fig:com}) or the explosion site (Figure~\ref{fig:origin}) as the origin of the multipole expansions. This finding implies that the small-scale (elliptical and mirror) asymmetries in the X-ray emission and the NS kicks may arise from or be affected by different processes. However, given the limitations of the geometric method (see Section~\ref{subsec:nsvel}), NS velocities estimated this way are not reliable (nine sources in our sample). Therefore, it is necessary to obtain more NS proper motion measurements before we can conclude robustly that small-scale asymmetries are uncorrelated with NS velocities. Additionally, we note that eight SNRs in our sample are interacting with molecular clouds. These interactions alter their X-ray morphologies and may produce their large power-ratios presented in Figure~\ref{fig:com}.  

We do find a correlation between the dipole angle and the direction of NS motion in the explosion-site analysis: five out of the six SNRs (CTB~109, Cas~A, Puppis~A, G292.0$+$1.8, and PKS~1209$-$51) with reliable explosion site constraints have dipole moments directed opposite to the NS kick (see Figure~\ref{fig:p1angle}). As discussed in Section~\ref{sec:results}, the X-ray emission in two of these SNRs (Cas~A, and G292.0$+$1.8) is ejecta-dominated. It is unknown if the emission from PKS~1209$-$51 arises from ejecta or CSM/ISM, and both CTB~109's and Puppis~A's interactions with molecular clouds may have affected their dipole angles. Consequently, it may be coincidental that these sources exhibit the $\sim$180$^\circ$ difference between dipole angle and NS direction. However, CTB~109 and Puppis~A do have distinct ejecta knots found in the opposite direction as their NSs' motion. It is unlikely that three out of the four interacting remnants by chance have emission directly opposed to their NSs' motion (consistent with the ejecta-dominated remnants) if the dipole moment direction is due to molecular cloud interaction alone. Thus, it is probable that the ejecta of these remnants contribute to their emission and dipole directions.

RCW~103 has a dipole angle in the same direction as the NS's motion. However, the X-ray morphology of RCW~103 may not reflect the ejecta distribution because of the bright emission produced by interaction with a molecular cloud. In addition, RCW~103 has no distinct, X-ray bright ejecta knots \citep{kari15} to influence the dipole direction, so the relation of the ejecta dipole moment to the emission dipole moment is unknown.

Given that both of the ejecta-dominated sources and three other SNRs with possible ejecta emission have NSs moving opposite to their emission, our results are consistent with NSs being kicked opposite to the bulk of the SN ejecta \citep{wongwathanarat13,muller16,janka17} rather than by anisotropic neutrino emission \citep{fryer06}.

Given the anti-correlation between NS direction and the position of the ejecta, we expect that there might be a relationship between the magnitude of the dipole power-ratios and the magnitude of the NS velocities. However, our small sample size prevents us from drawing a conclusion about the presence of such a trend. As shown in Figure~\ref{fig:origin}, the remnants affected by CSM/ISM interactions have systematically higher dipole asymmetries than the ejecta-dominated remnants. Only Cas~A and G292.0$+$1.8 are ejecta-dominated and two sources are not enough to draw a robust conclusion about the relationship between SNR ejecta and the magnitude of SNR velocities. In order to test robustly the relationship between NS velocities and the magnitude of SNR ejecta asymmetries, a larger sample of non-interacting SNRs with ejecta-dominated emission is needed.

We note that our bandpass is dominated by emission from O, Mg, and Si. Recent theoretical work predicts the distributions of O and Mg are only marginally correlated with NS motion, whereas heavier elements (Si and above) exhibit significant enhancement in the direction opposite to the NS kick \citep{wongwathanarat13}. These heavy metals are better tracers of explosion asymmetries; thus the dipole magnitude from only heavy element emission may show a stronger correlation with NS kick velocity than the emission in the 0.5--2.1 keV band. Future work can test this hypothesis, although only young SNRs have detectable radioactive Ti or are hot enough to produce Fe K$\alpha$ emission (the Fe L complex at $\sim$0.9--1.1~keV is not optimal for this analysis because it is blended with Ne lines at CCD energy resolution, and absorption attenuates the emission at these lower energies).

The uncertainties in NS velocities limit the inferences we can draw from the data.  The geometric method for assessing NS velocities requires several assumptions, as discussed in Section~\ref{subsec:nsvel}, making this method less reliable than other measurements. To get accurate transverse velocities, measurements of NS proper motions or of SNR filament expansions are needed. However, these studies require repeated, high spatial resolution images over $\gs$10 years. Among our sample, three SNRs (G11.2$-$0.3, G15.9$+$0.2, and W44) do not have NS proper motion measurements in the literature nor have long enough baselines in their multi-epoch {\it Chandra} ACIS observations to resolve the NS proper motion. Longer time baselines and the use of high spatial resolution imaging are crucial to measure NS positions and proper motions reliably.

\section{Conclusion} \label{sec:conc}

We have investigated the relationship between NS velocities and SNR soft X-ray morphologies as imaged by {\it ROSAT} and {\it Chandra} in a sample of 18 Galactic sources. NS velocities are estimated using either proper motion measurements or the geometric method, and SNR asymmetries are quantified using the power-ratio method. In our sample of SNRs with robust explosion site determinations, we found that five out of six (CTB~109, Cas~A, PKS~1209$-$51, G292.0$+$1.8, and Puppis~A) have NS motions directed opposite to the majority of their X-ray emission. The X-ray emission reflects the distribution of ejecta in two of these SNRs (Cas~A and G292.0$+$1.8). Two other SNRs (CTB~109 and Puppis~A) show evidence of ejecta emission in the direction of the dipole moment, but significant CSM/ISM contribution makes it unclear if their morphologies reflect the ejecta. It is unknown how interactions have affected the shape of PKS~1209$-$51. Although environment may influence the morphologies of these SNRs, it is unlikely that three out of the four interacting SNRs with well-constrained explosion sites have the same trend as the ejecta-dominated remnants unless ejecta contribute non-negligibly to the emission.

Contrary to the other remnants, in RCW~103 the dipole moment points in the same direction as the NS motion (toward the southeast). However, molecular cloud interactions have enhanced the emission there, influencing the direction of the dipole. As it is hard to disentangle the CSM emission from the ejecta emission in the rest of this SNR, it remains possible that the NS is moving opposite of ejecta.

As both Cas~A and G292.0$+$1.8 are ejecta-dominated, our results are consistent with predictions from models where the NS is kicked by ejecta asymmetries \citep{scheck06,wongwathanarat13,muller16,janka17}.

We did not find a correlation between the magnitude of SNR asymmetries and NS velocities. However, the absence of a trend may be due to the large uncertainties in NS velocity measurements and not being able to disentangle CSM/ISM and ejecta components sufficiently. In our sample, only two remnants are both ejecta-dominated and have precise NS velocities measured via proper motions. More X-ray observations of young SNRs and precise proper motion measurements of NSs are necessary to search for a robust trend in the relationship between NS velocities and SNR ejecta asymmetries.

Future work can help to elucidate the role of SN explosion and the surrounding environment in shaping SNRs. For example, SNRs' asymmetries may evolve with age, particularly as mature SNR morphologies reflect their environments (e.g., \citealt{gaensler98,west16}). In addition, explosion asymmetries may differ by nature or characteristics of the compact objects (e.g., neutron stars or black holes; magnetic field strength of the NSs). For example, \cite{duncan92} suggested that magnetars may be formed by progenitors with rapidly-rotating cores that lead to bipolar explosions. A correlation between SNR quadrupole power-ratio and NS magnetic field would support this scenario.

 \begin{deluxetable*}{lrrrrr}[b]
\tablecolumns{3}
\tablewidth{0pt} \tablecaption{\label{table:obslog}} 
\tablehead{\colhead{Source}  & \colhead{ObsID} &\colhead{Purpose\tablenotemark{a}} & \colhead{Exposure\tablenotemark{b}} & \colhead{Baseline\tablenotemark{c}} \\
\colhead{} & \colhead{} & \colhead{} & \colhead{(ks)} & \colhead{(yrs)}    }
\startdata
\cutinhead{{\it Chandra} Observations} 
G011.2$-$00.3 & 14830,14831,14832,15652,16323& PRM& 388& \\ 
 & 780,781,15652,16323 & NS & & 13\\
G015.9$+$00.2 & 5530,6288,6289,16766& PRM& 121&   \\
 & 5530,6288,6289,16766 & NS & & 10\\
Kes~73 & 6732,16950,17668,17692,17693& PRM& 120&  \\
 & 6732,17668,17692,17693 & NS & & 15\\
3C~391 & 2786& PRM& 60.9& \\
W44 & 5548,6312,11231&NS & & 10\\
3C~396 & 1988& PRM& 98.4& \\
Cas~A & 4634,4635,4636,4637,4638& PRM& 755& \\
IC~443 & 5531,13736&NS & & 7\\ 
MSH~11--62 & 14822,14823,14824,16497,16512,16541,16566& PRM& 358& \\
 & 2782,14822,14824 & NS & & 11 \\
G292.0$+$01.8 & 6677,6678,6679,8221,8447& PRM& 470 & \\
RCW~103 & 11823,12224,17460& PRM& 105& \\
 & 123,17460 & NS& & 16 \\
 \cutinhead{{\it ROSAT} Observations}
G018.9$-$01.1 & rp500040n00& PRM& 14 & \\
W44 & rp500019n00&  PRM& 7.5& \\
W51C & rp500152n00,rp500153a02,rp500153a01,rp500153n00&  PRM& 16.5& \\
CTB~109 & rp400314n00,rp500064n00& PRM& 40.6& \\
CTA~1 & rp500070n00,rp500072n00,rp500073n00& PRM& 24.5& \\
G189.1$+$03.0 & rp500020n00,rp500021n00,rp500022n00,rp500045a00,rp500045a01& PRM& 29.1& \\
Puppis~A & rp500055a01,rp500055n00,rp500056a01,rp500056n00,rp500057a01,rp500057n00& PRM& 17.2& \\
PKS~1209--51 & rp500239n00,rp500241n00,rp500242n00,rp500243n00&  PRM& 20.6& \\
MSH~15--56 & rp500010n00&  PRM& 3.78& 
\enddata
\tablenotetext{a}{PRM indicates the observations were merged and used to calculate the power-ratios. NS indicates that the observations were used to calculate a NS proper motion. For the latter, two epochs were used for each calculation: ObsID values below 10000 were merged for the first epoch, and ObsID values above 10000 were summed for the second epoch.}
\tablenotetext{b}{The summed exposure time in kiloseconds for all observations used to calculate the power-ratios. }
\tablenotetext{c}{Time in years between epochs of images used to calculate NS proper motion.}
\end{deluxetable*}

\begin{turnpage}

\begin{deluxetable*}{lcccccccccccl} 
\tablecolumns{7} \tablewidth{0pc} \tabletypesize{\scriptsize}
\setlength{\tabcolsep}{0.02in}
\renewcommand{\arraystretch}{1.0}
\tiny 
\tablecaption{Targets \label{table:SNRtable}}
\tablehead{
	\colhead{Source}          & \colhead{Alt.}      & \colhead{Size\tablenotemark{a}}     & \colhead{Age\tablenotemark{a}} 
    & \colhead{Distance\tablenotemark{a}}      & \colhead{NS}        & \colhead{NS}       & \colhead{NS}       
    & \colhead{NS}  & \colhead{NS}   &\colhead{Notes\tablenotemark{e}}   & \colhead{References}\\
	\colhead{}                & \colhead{Name}      & \colhead{(arcmin)} & (kyr)         
    & (kpc)                   & \colhead{Velocity}  & \colhead{Velocity} & \colhead{Direction\tablenotemark{c}} 
    & \colhead{Direction}     &\colhead{Method\tablenotemark{d}} &  & \colhead{}& \colhead{}\\
  	\colhead{}                & \colhead{}          & \colhead{}         &               
    &                         & \colhead{(km s$^{-1}$)} & \colhead{1-$\sigma$ Error\tablenotemark{b}} & \colhead{(Degrees)} 
    & \colhead{1-$\sigma$ Error} &\colhead{} &  & \colhead{}& \colhead{} } 
\startdata
G011.2$-$00.3 & SN AD386& 4.0 &1.65 &4.4  &  $<$108 & & & &geo& PWN & 1,2 \\ 
G015.9$+$00.2 & & 7.0 &2 &8.5  & 700  & & & &geo & & 3 \\
G018.9$-$01.1 & & 33 &5.3 &2 &  830  & & & & geo& PWN & 4,5\\
G027.4$+$00.0 & Kes 73 & 4.0 &2 &7.5  & 367  & 222& 78& 31& pm\tablenotemark{f}& AXP & 6,7\\ 
G031.9$+$00.0 & 3C 391 & 7.0 &17 &7.2 & 259  & & && geo &MC & 8,9,10,11 \\
G034.7$-$00.4\tablenotemark{g} & W44 & 35 &20 &3 & 370  &70& 160&  &mul & PWN; MM & 12,13,14 \\
G039.2$-$00.3 & 3C~396 & 8.0 & 3&6.2 & 344  & &&  &geo & PWN; MC & 15,16,17,18 \\
G049.2$-$00.7 & W51C & 30 & 30& 4.3  & 360  & &&  & geo& PWN; MC & 19,20,21 \\
G109.1$-$01.0 & CTB~109 & 28 &11 & 3.2  &  157 & 17& 253& 6& pm& AXP & 22,23,24 \\
G111.7$-$02.1 & Cas A & 5.0 & 0.33 &3.4  &  332  & 20& 170& 5&exp &CCO & 25,26 \\
G119.5$+$10.2 & CTA~1 & 90 &13 &1.4  &  450  & & & & geo&AXP & 27,28 \\
G189.1$+$03.0 & IC~443 & 45 & 10& 1.5&  160\tablenotemark{h}&  240& 270& 87& pm& MM; MC & 29 \\
G260.4$-$03.4 & Puppis A & 60 & 4.5& 2 &  437  & 72& 244& 11& pm& & 30,31 \\
G291.0$-$00.1 & MSH 11$-$62 & 15 &5.8 &5 & 303  & 130& 304& 22&pm\tablenotemark{f} & PWN & 32,33 \\
G292.0$+$01.8 & MSH 11-54 & 12 &3 &6.3 & 462  & 31& 128& 11&exp & PWN & 34,35,36 \\
G296.5$+$10.0 &PKS~1209$-$51 & 90 &7 & 2& 142  & 66& 307& 39& pm& & 37,38,39 \\
G326.3$-$01.8 & MSH 15$-$56 & 38 &16.5 &4.1 &   410 & & & & geo& & 40,41,42 \\
G332.4$-$00.4 & RCW~103 & 10 & 2& 3.3&  169  & 51& 163& 8& pm\tablenotemark{f}& SGR; MC& 7,43 \\
\enddata
\tablenotetext{a}{The values used in our analysis. The errors vary from source to source and can be found in the references.}
\tablenotetext{b}{We do not include error bars for or directions of the velocities calculated using the geometric method, as they do not reflect the uncertainty in attributing the geometric center as the explosion site.}
\tablenotetext{c}{Direction is in the format where 0$^\circ$ points North, 90$^\circ$ points East, etc.}
\tablenotetext{d}{``pm" denotes proper motion method, ``geo" denotes geometric estimate, ``exp" denotes expansion center via filament motion, and ``mul" denotes multiple converging estimates.}
\tablenotetext{e}{Notes include whether the source has a pulsar wind nebula (PWN), an anomalous X-ray pulsar (AXP), a soft gamma-ray repeater (SGR), or a Central Compact Object (CCO), whether the source is adjacent to or is interacting with an H {\sc ii} region or molecular cloud (MC), and whether the source is considered a ``mixed-morphology'' SNR.}
\tablenotetext{f}{The proper motion was calculated in this work using {\it Chandra} ACIS observations.}
\tablenotetext{g}{The direction of motion of the NS in W44 is not reported in the literature.}
\tablenotetext{h}{The proper motion of the NS in IC~443 was calculated in \cite{swartz15} using {\it Chandra} ACIS observations over a 7--year baseline. It is consistent with zero at the 1-$\sigma$ level and has large directional errors, so we use the geometric estimate of 615 km s$^{-1}$ in our analysis.}
\tablecomments{ References are as follows \newline
[1] \cite{kaspi01}
[2] \cite{dean08}
[3] \cite{reynolds06}
[4] \cite{harrus04} 
[5] \cite{tullmann10} 
[6] \cite{kuiper04}
[7] \cite{lopez11} \newline
[8] \cite{chen04}
[9] \cite{kawasaki05}
[10] \cite{su14}
[11] \cite{sato14}
[12] \cite{harrus97}
[13] \cite{shelton04}
[14] \cite{frail96} \newline
[15] \cite{harrus99} 
[16] \cite{safi05}
[17] \cite{olbert03}
[18] \cite{su11}
[19] \cite{koo95}
[20] \cite{koo05}
[21] \cite{tian13} \newline
[22] \cite{sandy01}
[23] \cite{tendulkar13}
[24] \cite{kothes12}
[25] \cite{hwang04}
[26] \cite{thorstensen01}
[27] \cite{halpern04} \newline
[28] \cite{slane04} 
[29] \cite{swartz15}
[30] \cite{becker12}
[31] \cite{reynoso17}
[32] \cite{harrus98}
[33] \cite{slane12} 
[34] \cite{vink04} \newline
[35] \cite{park07}
[36] \cite{winkler09} 
[37] \cite{vasisht97}
[38] \cite{zavlin00}
[39] \cite{halpern15}
[40] \cite{kassim93} \newline
[41] \cite{temim13}
[42] \cite{sun99}
[43] \cite{dai16}}
\end{deluxetable*}  
\end{turnpage}

\acknowledgements

We thank the referee for their helpful feedback which improved the quality of the manuscript. We acknowledge helpful discussions with Dr. Christopher Kochanek and Diego Godoy-Rivera. This work is supported through NSF Astronomy \& Astrophysics Grant AST--1517021 and The Ohio State University's Dean's Distinguished University Fellowship. This research made use of the {\it Chandra} and {\it ROSAT} Data Analysis Software ({\sc ciao} and {\sc ftools}).

\noindent
{\it Facilities}: {\it Chandra}, {\it ROSAT}
\nocite{*}
\bibliographystyle{apj}
\bibliography{NSkick}

\end{document}